%% file: ms.tex
\documentclass[usenatbib]{raa}            
\usepackage{graphicx,times}             
\usepackage{epstopdf}
\usepackage{lscape}
\usepackage{pdflscape}
\usepackage{colortbl}
\usepackage{amsmath}
\usepackage{textcomp}
\usepackage{multicol}
\usepackage{longtable}
\usepackage[figuresright]{rotating}
\usepackage{natbib}
\usepackage[colorlinks,  linkcolor=cyan,  anchorcolor=black,  citecolor=cyan]{hyperref}

\begin{document}
\newcommand{\ha}{\hbox{H$\alpha$}}
\newcommand{\hb}{\hbox{H$\beta$}}
\newcommand{\hd}{\hbox{H$\delta$}}
\newcommand{\hg}{\hbox{H$\gamma$}}
\newcommand{\he}{\hbox{H$\epsilon$}}
\newcommand{\oi}{\hbox{[O\,{\sc i}]}}
\newcommand{\oii}{\hbox{[O\,{\sc ii}]}}
\newcommand{\nii}{\hbox{[N\,{\sc ii}]}}
\newcommand{\sii}{\hbox{[S\,{\sc ii}]}}
\newcommand{\oiii}{\hbox{[O\,{\sc iii}]}}
\newcommand{\neiii}{\hbox{[Ne\,{\sc iii}]}}
\newcommand{\hei}{\hbox{[He\,{\sc i}]}}
\newcommand{\heii}{\hbox{[He\,{\sc ii}]}}
\newcommand{\dfn}{\hbox{$D_{\rm n}$(4000)}}
\newcommand{\hda}{\hbox{H$\delta_{\rm A}$}}
\newcommand{\zoh}{\hbox{$12\, +\, {\rm log(O/H)}$}}
\newcommand{\znh}{\hbox{$12\, +\, {\rm log(N/H)}$}}
\newcommand{\etal}{\hbox{et\thinspace al.\ }}
\newcommand{\mum} {\hbox{$\mu{\rm m}$}}
\newcommand{\Zsun}{\hbox{${\rm Z}_{\sun}$}}
\newcommand{\Msun}{\hbox{$M_{\sun}$}}
\newcommand{\hi}{\hbox{H\,{\sc i}}}
\newcommand{\hn}{\hbox{H\,{n}}}
\newcommand{\nodata}{$\cdots$}

\title{A sample of metal-poor galaxies identified from the LAMOST spectral survey}

\volnopage{Vol.  0 (200x) No.  0,   000--000}      
\setcounter{page}{1}          

\author{
	Yulong Gao\inst{1,2},  
	Jianhui Lian\inst{1,2},  
        Xu Kong\inst{1,2},
        Zesen Lin\inst{1,2},
        Ning Hu\inst{1,2},
        Haiyang Liu\inst{1,2},
        Enci Wang\inst{1,2},
        Zihuang Cao\inst{3},
        Yonghui Hou\inst{4},
	Yuefei Wang\inst{4},
	\and 
	Yong Zhang\inst{4}
     }
	\institute{CAS Key Laboratory for Research in Galaxies and Cosmology, 
        Department of Astronomy, University of Science and Technology of China, 
        Hefei 230026, China; {~~~~\it E-mail: ylgao@mail.ustc.edu.cn; xkong@ustc.edu.cn}\\
   \and School of Astronomy and Space Science, University of Science and
        Technology of China, Hefei 230026, China \\
   \and Key Laboratory of Optical Astronomy, National Astronomical Observatories, 
        Chinese Academy of Sciences, Beijing 100012, China \\
   \and Nanjing Institute of Astronomical Optics $\&$ Technology, National Astronomical 
        Observatories, Chinese Academy of Sciences, Nanjing 210042, China
   }

   \date{Received~~2009 month day; accepted~~2009~~month day}

\abstract{We present a sample of  48 metal-poor galaxies at $ z < 0.14$ selected 
from  92,510 galaxies in the LAMOST survey.  
These galaxies are 
identified for their detection of the auroral emission line \oiii$\lambda$4363
above $3\sigma$ level,  which  allows a direct measurement of the electron temperature
and the oxygen abundance.  
The emission line fluxes are corrected for internal dust extinction using 
Balmer decrement method. With electron temperature derived from 
\oiii$\lambda\lambda4959,5007/\oiii\lambda4363$ and electron density from 
$\sii\lambda6731/\sii\lambda6717$, we obtain the oxygen abundances in
our sample which range from $\zoh= 7.63$ (0.09 $\Zsun$) to $8.46$ (0.6 
$\Zsun$).
 We find an extremely metal-poor galaxy with $\zoh=7.63 \pm 0.01$.
With multiband photometric data from FUV to NIR and $\ha$ measurements, we also 
determine 
the stellar masses and star formation rates, based on the spectral energy 
distribution fitting 
and  $\ha$ luminosity,
respectively. We find that our galaxies have low and intermediate stellar masses with
$\rm 6.39 \le log(M/M_{\sun})\le 9.27$, and high star formation rates (SFRs) 
with
$\rm -2.18 \le log(SFR/M_{\sun} yr^{-1}) \le 1.95$. We also find that the metallicities 
of our galaxies are consistent with the local $T_e$-based
mass-metallicity relation,  
while the scatter is about 0.28 dex. Additionally, 
assuming the coefficient of $\rm \alpha =0.66$,
we find most of our galaxies follow the local mass-metallicity-SFR relation, 
while a scatter about 
0.24 dex exists, suggesting the mass-metallicity relation is weakly dependent 
on SFR for those metal-poor galaxies.
}
   \authorrunning{Y.L. Gao et al. }      
   \titlerunning{A sample of metal-poor galaxies in the LAMOST spectral survey} 

   \maketitle
\section{Introduction}          
\label{sect:intro}

  Metal-poor galaxies are less chemically evolved galaxies and provide ideal laboratory 
  for investigating galaxy properties in extreme 
  condition \citep{2005AA...437..849S, 2016ApJ...819...73L}.  Among them,
  extremely metal-poor galaxies (hereafter XMPGs), defined  by their low oxygen 
  abundance with $\zoh \leq 7.65$ 
  \citep{2003ApJ...593L..73K, 2007AA...464..859P, 2005MNRAS.361...34D}, are the
  most promising young galaxy candidates in the local universe \citep{2004ApJ...616..768I}. 
  These XMPGs are suspected to be primeval galaxies that are undergoing their first major
  mass assembly at the observed redshift \citep{2003ApJ...593L..73K}. Studying these extreme 
  objects can improve our understanding about the early stages of galaxy assembly.  
  
The determination for abundance of elements are considered more reliable if the electron
temperature $T_e$ can be measured directly, because the metallicity is anti-correlated with
the electron temperature. The electron temperature can be obtained  using the auroral line
ratios, such as \oiii$\lambda4363/$\oiii$\lambda\lambda4959,5007$. 
This technique is often called the $T_e$  method \citep{1984ASSL..112.....A}. 
   However, galaxies with metallicities derived
from the $T_e$ method with $\oiii\lambda4363$ detections above 3$\sigma$ are extremely rare. 
To date, only  about 174 such objects  has been 
found \citep{2014ApJ...780..122L, 2016ApJ...828...67L}.

In order to enlarge  the sample  of  metal-poor galaxies  with 
\oiii$\lambda$4363 detection, we carry out a systematic search for such objects 
in the LAMOST Data Release (DR3, DR4 Q1 and Q2).  In Section \ref{sect:Sample}, we present
our approach for detecting and measuring nebular emission lines, and the selection 
criteria used to identify the metal-poor galaxies.  We describe the determination
for the dust attenuation properties and the gas-phase oxygen abundances in Section
\ref{metallicity determination}.
In Section \ref{stellar}, we describe the methods for estimating 
stellar masses and SFRs, and then compare the 
mass-metallicity relation (MZR) and mass-metallicity-SFR relation
 with local $T_e$-based relation  in Section \ref{mz_fmr}. In addition, we 
discuss our results in the context of other studies in Section 
\ref{discussion}. 
Finally, we summarize our main conclusions in Section \ref{summary}.    

Throughout this paper, we adopt a flat cosmology with $\Omega_\Lambda=0.7$,  
$\Omega _M=0.3$,  and $H_0=70$ km s$^{-1}$ Mpc$^{-1}$ to determine 
distance-dependent measurements.  For reference,  we adopt 
$\zoh_{\sun} = 8.69$ \citep{2001ApJ...556L..63A} for metallicity 
measurements quoted against the solar value, $\Zsun$. 
\section{Sample selection}
\label{sect:Sample}
\subsection{LAMOST dataset}
\label{sect:2.1}
The Large Sky Area Multi--Object Fiber Spectroscopic Telescope (LAMOST, also called the 
Guo Shou Jing Telescope) is a special reflecting Schmidt telescope with an effective 
aperture of 4 m and a field of view (FoV) of 5$^{\circ}$ \citep{Wang:96, 1009-9271-4-1-1, 2012RAA....12..723Z, 2012RAA....12.1197C, 1674-4527-12-9-004}. It is equipped with 4000 fibers, 
covering a wavelength range of 3800 $-$ 9000 $\mathrm{\AA}$ \citep{2015RAA....15.1095L} at a resolving power R $\approx$ 1800. The LAMOST Data Release 3 (DR3), Data Release 4 (DR4)  Q1 and  Q2, based
on the past survey from October 2011 to  February 2016, contain about  three million spectra with limiting magnitude of $r \approx$ 18.5 mag. The LAMOST 1D pipeline classifies spectroscopic targets as galaxies,
 stars, and QSOs by matching against observed SDSS spectral templates, see 
\cite{2015RAA....15.1095L} for detail. 92,510 objects from the LAMOST catalog are spectroscopically classified as galaxies (`OBJTYPE' = `GALAXY').

\subsection{Emission-line fluxes determination}
\label{sect:2. 2}

According to \cite{2012RAA....12..453S} and \cite{2015RAA....15.1095L}, LAMOST
1D spectra are extracted from the CCD images used by the LAMOST 2D pipeline. The
wavelength calibration of each spectrum is accomplished by using arc lamp spectra
lines, with an average calibration error less than 0.02$\rm \AA$.  The accuracy 
of the relative flux calibration
of LAMOST is above 90$\%$.

Assuming unimodal gaussian line profile, we obtain fluxes of strong 
emission lines
such as $\oii\lambda3727$, $\oiii\lambda4363$,  $\hb$,  $\oiii\lambda\lambda4959,5007$, $\ha$
and $\sii\lambda\lambda6717,6731$ by fitting their line profile using the 
IDL package \textsc{mpfit} \citep{2009ATel.2258....1M}. The expected location 
of emission lines are based on a priori redshift determined by  the $\oiii$ line. 
In addition, to estimate the signal-to-noise ratios (S/N) of emission lines, 
we follow the 
calculation
method in \cite{2014ApJ...780..122L}.

\subsection{Sample selection}
\label{sect:2.3}

Among all the galaxies from the LAMOST ExtraGAlactic Surveys (LEGAS), we first 
select
a subsample of metal-poor galaxies
with emission line flux ratios $\nii\lambda6583/\ha \le 0.1$, which consists of 665 galaxies.
Among them, we identify  237 objects with  $\oiii\lambda4363$
detection at $\ge 3 \sigma$.  
We inspect these 237 objects visually, and find 73 of them are false 
detections. We also exclude 
 115 objects  
that are $\rm \hii$ regions in nearby large galaxies  using optical 
images with SDSS 
DR12 skyserver \footnote{http://skyserver.sdss.org/dr12/en/tools/chart/listinfo.aspx}. 
 Finally, we check the right ascension and declination of  the remaining sources, and 
note that  1 object was observed  twice by LAMOST.  We 
keep the observation that has better spectral quality.

\begin{figure}[h]
\center
\includegraphics[width=0.6\textwidth]{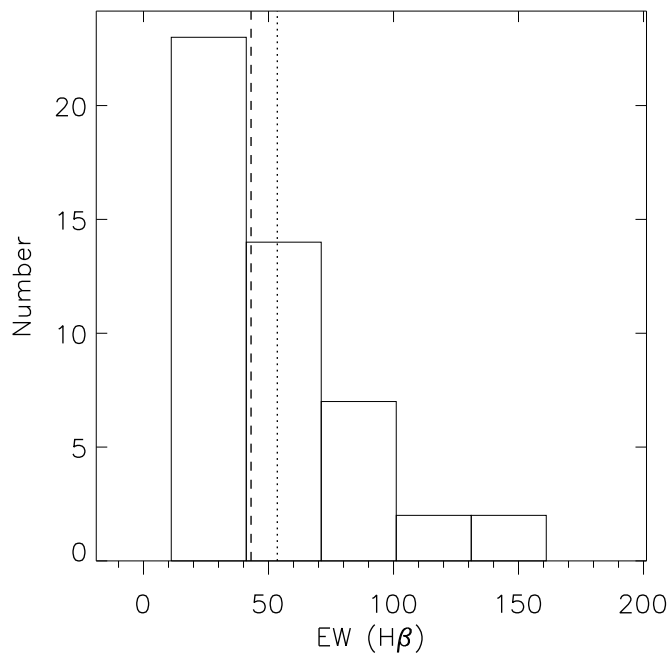}
\caption{The distribution of $\hb$ equivalent widths in our sample. 
The dashed line and dotted line represent the median and average equivalent 
widths of 
42.9 $\rm \mathrm{\AA}$ and 53.5 $\rm \mathrm{\AA}$, respectively.} 
\label{hist_hb}
\end{figure}

As a consequence, our final sample consists of  48 galaxies, making up
only $\rm 0.05\%$ of all the LAMOST galaxies until
DR4 Q2; this fraction is nearly the same as SDSS \citep{2014ApJ...780..122L}. 
The median S/N of $\oiii\lambda4363$ is  6.1. 
We obtain the $\hb$ equivalent width (EW) by dividing the $\hb$ flux by 
continuum spectral flux
intensity, which is assumed as the average value of observed flux intensities 
within
50$\rm \AA$ wide component around the $\hb$ line. 
All of the galaxy spectra in our sample show strong emission lines with a 
median (average) EW of $\hb$ of  42.9 (53.5) $\mathrm{\AA}$. 
The EW distribution of  $\hb$ is shown in Figure \ref{hist_hb}.

\begin{figure}
\centering
\includegraphics[width=0.7\textwidth]{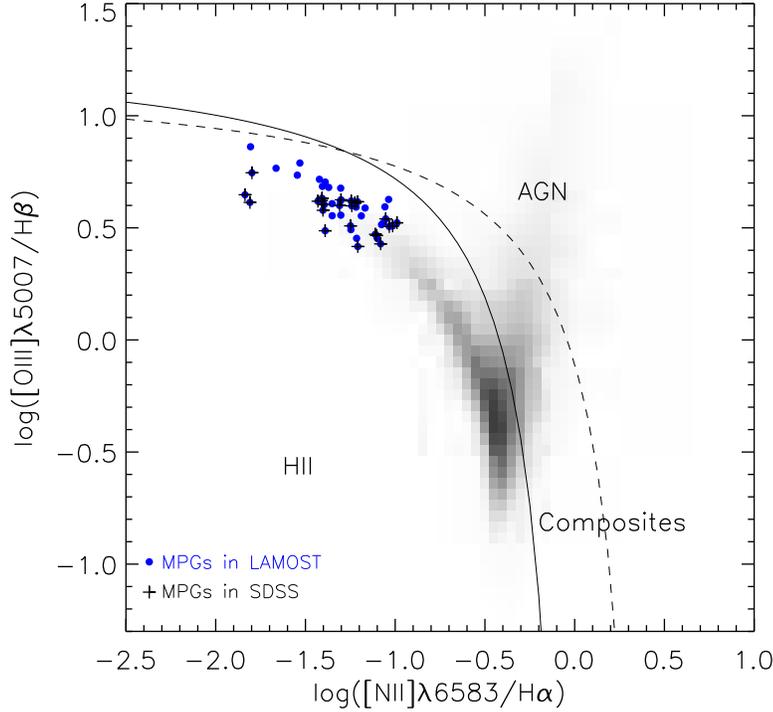}
\caption{ BPT diagram for our metal-poor galaxy (MPG) sample and all the 
LAMOST galaxies.  
The blue 
dots represent our final sample galaxies.    
The black crosses represent these objects in our final sample that have also 
been
spectrally detected with SDSS. The
grayscale 2D histogram shows the number density of LAMOST galaxies.
The solid and dashed lines are the demarcation curves between SFGs and AGNs 
defined by \cite{2003MNRAS.341...33K} and \cite{2001ApJ...556..121K}, 
respectively.}
\label{bpt}
\end{figure}

 To exclude  possible AGN contamination, we use the BPT diagram
 \citep[see][]{1981PASP...93..817B,1987ApJS...63..295V,
 2001ApJ...556..121K,2003MNRAS.341...33K}. 
Figure \ref{bpt} shows the distribution of our sample  in the BPT diagram.
The grayscale 2D histogram shows the number density of LAMOST galaxies.
The blue dots represent 48 galaxies in our final sample, and the black crosses 
represent 24 galaxies in our final sample that have also been
spectrally detected by SDSS. Among these 24 galaxies detected with SDSS, 
19 galaxy spectra also have the $\oiii\lambda4363$ detections above 3$\sigma$; 
 we will compare these 19 galaxy spectra from LAMOST and SDSS in Section 
\ref{comp_spec}. 
The solid and dashed lines are the demarcation curves between SFGs and AGNs 
derived by 
 \cite{2003MNRAS.341...33K} and \cite{2001ApJ...556..121K}. 
 Galaxies located between the two lines are usually classified as composite
 objects, which may host a mixture of star formation and AGN.  
 It can be seen that all of the galaxies in our final sample are located in the 
 star-forming region; however, this is unsurprising since we initially selected 
sources
 with $\nii\lambda6583/\ha \le 0.1$. 
 
\begin{figure}[h]
\center
\includegraphics[width=1.0\textwidth]{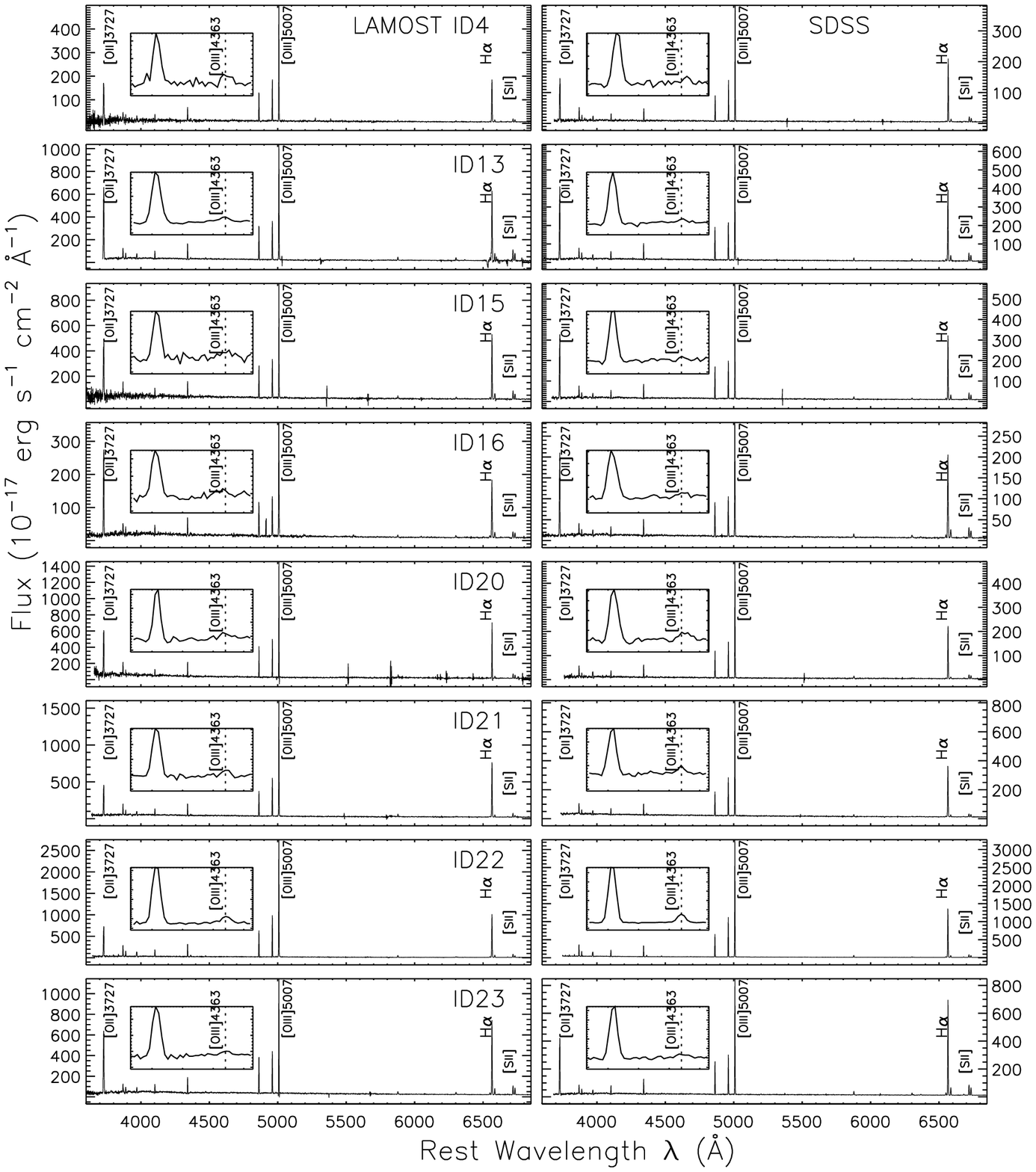}
\caption{Example spectra of objects in our sample that 
have both been spectrally detected with LAMOST and SDSS. 
For every object, the left panel shows LAMOST spectrum, while the right panel
shows SDSS spectrum. Inserted panels show the zoomed in spectra adjacent 
to the \oiii$\lambda4363$ lines.}  
\label{spec}
\end{figure}

Figure \ref{spec} shows example spectra for eight galaxies
in our sample that have both been
spectrally detected by LAMOST and SDSS. For each object, the left panel shows 
the 
LAMOST spectrum, while the right panel shows the SDSS spectrum. All of these 
spectra show strong emission lines such as $\oii\lambda3727$ (for all LAMOST 
spectra and
part of SDSS spectra),
$\hb$, $\oiii\lambda\lambda4959,5007$, $\ha$ and $\sii\lambda\lambda6717,6731$.
The inserted panels show the zoomed spectra adjacent to $\oiii\lambda4363$ 
lines. It can be seen 
that the weak \oiii$\lambda4363$ lines are all  detected in the spectra of these 
galaxies.
 
\section{Metallicity determination}
\label{metallicity determination}
\subsection{Dust attenuation correction}

We correct the emission-line fluxes for internal dust attenuation using the 
Balmer 
decrement measurements, which estimate the dust extinction by inspecting the 
change of Balmer
line ratio, such as $\ha/\hb$, from intrinsic value.
Generally, the underlying stellar absorption in the Balmer lines should be well 
determined 
to obtain a reliable emission line measurement \citep{1674-4527-16-3-006}. 
 In this work, we first subtract the underlying stellar continuum 
and stellar absorption for each spectrum using the \textsc{starlight} spectral 
synthesis
code \citep{2005MNRAS.358..363C}. 
We assume the intrinsic flux ratio 
of $(\ha/\hb)_0=2.86$ \citep{1987MNRAS.224..801H}  under  Case B recombination 
and use the \cite{2000ApJ...533..682C}  reddening formalism to derive the color 
excesses
$E(B-V)$, and then  correct the emission line fluxes. In addition, we manually 
set the
color excesses $E(B-V)$ to zero when the $\ha/\hb$ ratios are less than 2.86.
 
The resulting reddening-corrected emission line fluxes relative to $\hb$ and 
color 
excesses are listed in Table 1. As showed in panel $a$ of Figure 
\ref{plot_hist}, 
the measured dust extinction are very low with an average $E(B-V)$ value as 
0.03 mag.

\subsection{Metallicity calculation}

With significant detection of \oiii$\lambda4363$, we can determine the 
metallicity using 
the so-called $T_e$ method. 
In this work, we use the python package \textsc{pyneb}  
\footnote{http://www.iac.es/protecto/PyNeb/}
\citep{2015AA...573A..42L} to calculate the electron densities ($n_e$) 
and electron temperatures ($T_e$), which is evolved from the IRAF  nebular 
package
\citep{1995PASP..107..896S,1998ASPC..145..192S}.
\cite{2013ApJS..207...21N} have demonstrated that
the electron temperatures would be overestimated, and thus the oxygen
abundances would be underestimated when using older collision strength
data and approximate temperature calibration methods from  
\cite{2006AA...448..955I}.
  Therefore,
we need to set the atomic recombination data and atomic collision strength data 
before
the calculation. We adopt the atomic recombination data  of
\cite{2004ADNDT..87....1F} for $\rm O^+$, $\rm O^{++}$,  and
\cite{2010ApJS..188...32T} for $\rm S^+$. 
For collision strength data, we adopt those from 
\cite{2009MNRAS.397..903K} for $\rm O^+$,
\cite{2014MNRAS.441.3028S} for $\rm O^{++}$, and \cite{2010ApJS..188...32T} for 
$\rm S^+$.
 
As encouraged in \cite{2015AA...573A..42L}, we use a cross-converging method to 
calculate 
the electron temperatures of $\rm O^{++}$ regions ($T_e(\oiii)$) and electron 
densities
($n_e$) with ratios of
 $\oiii\lambda4363/\oiii\lambda\lambda4959,5007$ \citep{2013ApJS..207...21N} 
 and $\sii\lambda6717/\sii\lambda6731$ \citep{2010ApJS..188...32T}.
  Once the electron temperatures $T_e(\oiii)$
 and densities $n_e$ are determined, we can obtain the ionic oxygen $\rm O^{++}$ 
abundances 
 using the $\oiii\lambda\lambda4959,5007/\hb$ ratios with the relation 
derived from  \cite{2006AA...448..955I}.  
In order to derive the electron temperatures $T_e(\oii)$ of $\rm O^+$ regions, 
we follow an
iterative method used in \cite{2014ApJ...786..155N},
\begin{equation}
 T_e(\oii) = T_e(\oiii) \times \rm (3.0794-0.086924Z-0.1053Z^2+0.010225Z^3),
\label{te_oii_zoh}
\end{equation}
where $\rm Z$ is the total oxygen abundance, $\rm 12+log(O/H)$.
The temperature $T_e(\oii)$ and abundance $\rm Z$ will converge within five 
iterations,
starting by using the $\rm O^{++}$ abundance as the total oxygen abundance. 
 Here, the $\rm O^{+}$ abundance is determined from $T_e(\oii)$ and the 
$\oii\lambda3727/\hb$ ratio using the \cite{2006AA...448..955I} relation. 
Using other methods from \cite{1992AJ....103.1330G} or 
\cite{2012MNRAS.426.2630L},
we will get a higher $T_e(\oii)$ about 0.02 dex and a lower metallicity 
about 0.05 dex  given a $T_e(\oiii)$. Similarly, we will get a higher 
$T_e(\oii)$  about 0.03 dex and thus get a lower metallicity about 0.05 dex 
when adopt the standard two-zone temperature model from 
\cite{2006AA...448..955I} and \cite{2013ApJ...765..140A}. 
In our final sample, 31 galaxies also have the 
$\oii\lambda\lambda7320,7330$  detections above 3$\sigma$.   Using the 
$\oii\lambda3727/\oii\lambda\lambda7320,7330$ ratios to derive $T_e(\oii)$, 
we will get lower $T_e(\oii)$ and higher 
metallicities, the differences between these average values on $T_e(\oii)$ 
and metallicities are about 360 $\rm K$ and 0.06 dex, respectively.

To estimate the uncertainties of electron temperatures, electron densities and 
oxygen
abundances, we repeat the calculation 2000 times. For every object, we produce a 
series
of fluxes for each emission line with a Gaussian distribution, assuming its 
average is 
the measured line flux and standard deviation is the measured error. Then, our 
final 
temperature, density and oxygen abundance are deemed to be the median values of 
these
2000 calculations, and the corresponding errors are estimated as the half of 
$16\%\--84\%$ range of their distributions.     
We list the final electron temperatures, electron densities and oxygen 
abundances
of our sample in Table 1. 

\begin{figure}[h]
\center
\includegraphics[width=0.8\textwidth]{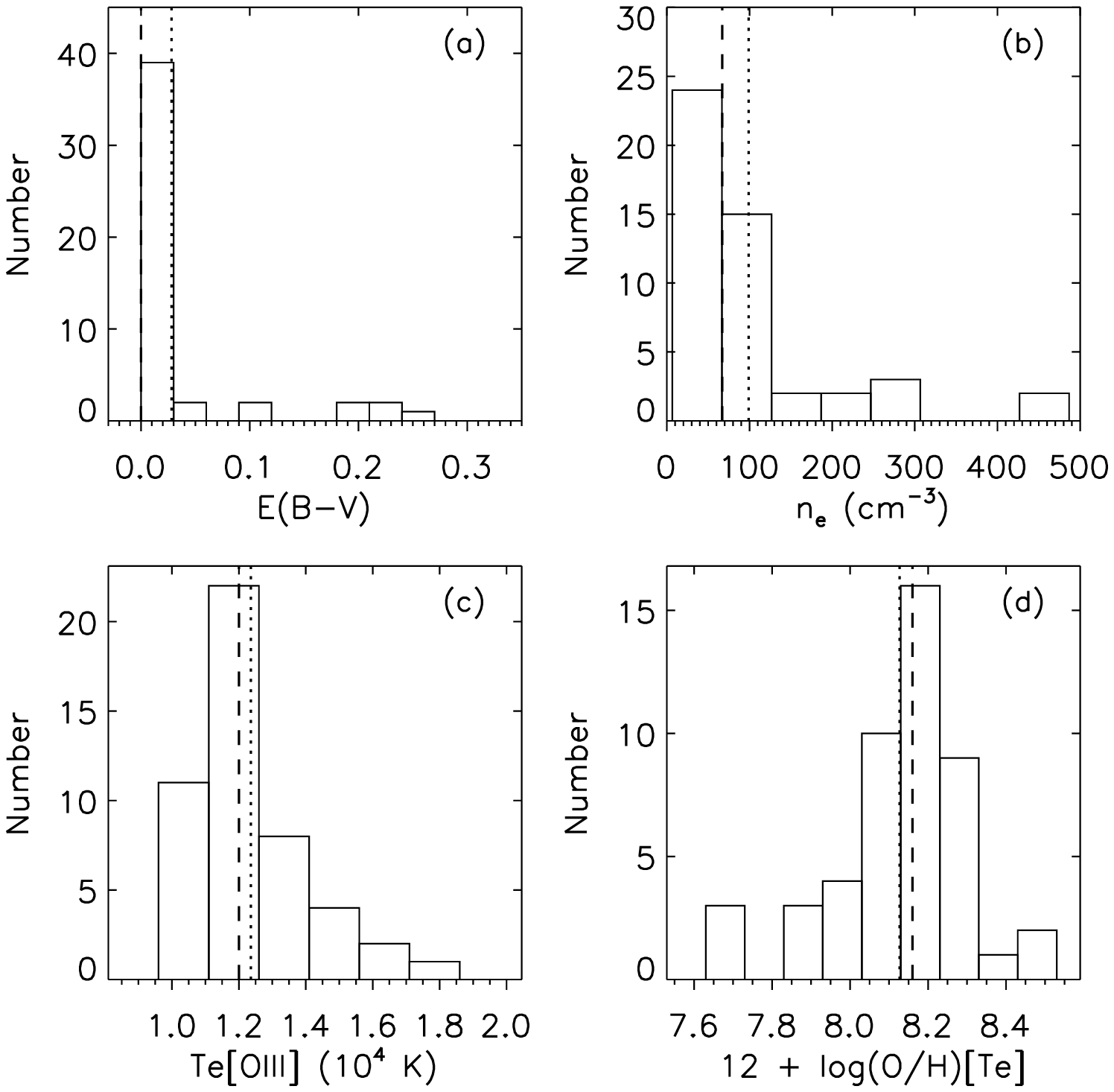}
\caption{The distributions of color excesses, electron densities, electron 
temperatures
and oxygen abundances of our sample galaxies. The dashed lines represent the 
median values of these 
parameters, 0.00 mag, 67.0 $\rm cm^{-3}$, 1.20 $\rm \times 10^4 \ K$ and 8.16, 
respectively. The dotted 
lines represent the average values 0.03 mag, 98.9 $\rm cm^{-3}$, 1.24
$\rm \times 10^4 \ K$ and 8.13, respectively.}
\label{plot_hist}
\end{figure}

Figure \ref{plot_hist} shows the distributions of color excesses, electron 
densities,
electron temperatures and oxygen abundances.  The electron densities and 
temperatures
in our sample range from $\rm 6.7$ to $\rm 475.0 \ cm^{-3}$ 
and $\rm (0.96\--1.82)\times 10^4 \ K$, with median values of $\rm 67.0 \ 
cm^{-3}$ and
$\rm 1.21\times 10^4 \ K$, respectively.  Their oxygen abundances range from 
7.63 to 8.46, with a median of 8.16. The only  XMPG found in our sample is 
ID26 with
$\rm \zoh=7.63\pm0.01$, which has already been  found by 
\cite{2012AA...546A.122I} with 
$\rm \zoh=7.64\pm0.01$. Interestingly, galaxy ID33, also named as RC2 
A1228+12, was 
regarded as an XMPG in \cite{2000AARv..10....1K} and \cite{2014MNRAS.441.2346B}, 
but the metallicity 
$\rm \zoh=7.70\pm0.01$ indicates it is not an XMPG. This judgement was
also supported by \cite{2002AA...389..779P} and \cite{2012MNRAS.427.1229I} with 
metallicity
measurements of $\rm 7.73\pm0.06$ and $\rm 7.70\pm0.01$, respectively. 

\section{stellar masses and star formation rates}
\label{stellar}

\subsection{Stellar masses}
\label{stellar masses}
To determine the galactic stellar masses of our sample galaxies, we use the IDL 
code library 
\textsc{fast} developed by \cite{2009ApJ...700..221K} to perform the spectral 
energy distribution
(SED) fitting. \textsc{fast} compares the photometry measurements with stellar 
population 
synthesis models, based on the minimum $\chi ^2$ template-fitting procedure, to 
determine 
mass-to-light ratios, which can be used to estimate the stellar masses of 
galaxies.  
We use the stellar templates of \cite{2003MNRAS.344.1000B} and a
\cite{2003PASP..115..763C} initial 
mass function (IMF) to synthesize magnitudes. These models span four 
metallicities (0.004, 0.008, 0.02, 0.05 $\Zsun$) and an exponentially decreasing 
star
formation models (SFR $\propto \rm e^{-t/\tau}$) with a step $\Delta log(\tau) = 
0.1$ 
from $6.6 \leq log(\tau) \leq 10.8$. We assume the dust attenuation law from
\cite{2000ApJ...533..682C} allowing $E(B-V)$ to vary from 0.0 to 2.0 and stellar 
population
ages ranging from 0 to 100 Gyr. To determine the uncertainties of stellar 
masses, we use the Monte Carlo simulations and define the number of simulations 
as 1000. We choose the confidence interval as 68$\%$. 

Photometric measurements are collected from various survey catalogue. We adopt 
values 
of \textsc{modelmag} magnitudes of $u, g, r, i$, and $z$ bands from SDSS DR12 
photometry catalogue
\citep{2004AJ....128..502A, 2015ApJS..219...12A}, 
magnitudes of $J, H$, and $K_s$ bands from 2MASS All-Sky Point Source catalog 
(PSC)
and 2MASS All-Sky
Extended Source Catalog (XSC) \citep{2006AJ....131.1163S}, magnitudes of $W1$ 
(3.4 $\mu m$) and $W2$ (4.6
$\mu m$) from All WISE Source Catalog \citep{2010AJ....140.1868W}, and 
magnitudes of FUV 
and NUV from $GALEX$ GR6/7 Data Release Catalog \citep{2014AdSpR..53..900B}.
However, not all of our sample galaxies have these photometric measurements. For 
example,
 45 
galaxies have FUV photometry, while 3 galaxies are not
    located in $GALEX$ surveyed areas. For these three galaxies, we 
just use their  magnitudes from $u$ band to $W2$ band to 
perform the SED fitting. 
We find our sample galaxies spanning three orders with 
$\rm 6.39 \leq log(M/M_{\sun}) \leq 9.27$. We should note that we do not 
make the point spread function (PSF) matching for our photometric data using 
same observation aperture, which may lead to some uncertainties in stellar 
mass measurements. The average and median values on stellar mass 
measurement uncertainties are 0.14 dex and 0.12 dex, respectively. However, 
comparing our results with total stellar masses in 
MPA-JHU 
catalog \citep{2003MNRAS.341...33K, 2004MNRAS.351.1151B} for these galaxies 
included in MPA-JHU catalog, we find the differences of average and median 
values are about 0.1 dex and 0.03 dex, respectively. 

\input{tab1.tex}

\label{table1}

\subsection{Star formation rates}
\label{SFR}

In this work, we use the $\ha$ emission line luminosities to 
determine dust-corrected SFRs, assuming a \cite{2003PASP..115..763C} IMF and solar metallicity. The SFR can be calculated from $\ha$ luminosity as:
\begin{equation}
{\rm SFR(M_{\sun} yr^{-1})} = R \ {\rm \times \ L(erg \ s^{-1})},
\end{equation}
where $ R = 4.4 \ \times \ {\rm 10^{-42}}$. However, the latest work of 
\cite{2016ApJS..226....5L} demonstrated that the above parameter $R$
would overestimate the SFR at lower metallicities, and gave the
metallicity-dependent parameter $R$ as: 
 \begin{equation}
 {\rm log}(R) = {\rm log(\frac{SFR}{L(\ha)})} = -41.34 + 0.39y + 0.127y^2,
 \end{equation}
where $y = \rm log(O/H) + 3.31$. 
Above all, the final stellar masses and SFRs in our sample are listed in Table 1.

\section{The Mass--Metallicity and Mass--Metallicity--SFR Relations}
\label{mz_fmr}

\subsection{The mass-metallicity relation}
\label{mz-rela}
\begin{figure}[h]
\center
\includegraphics[width=1.0\textwidth]{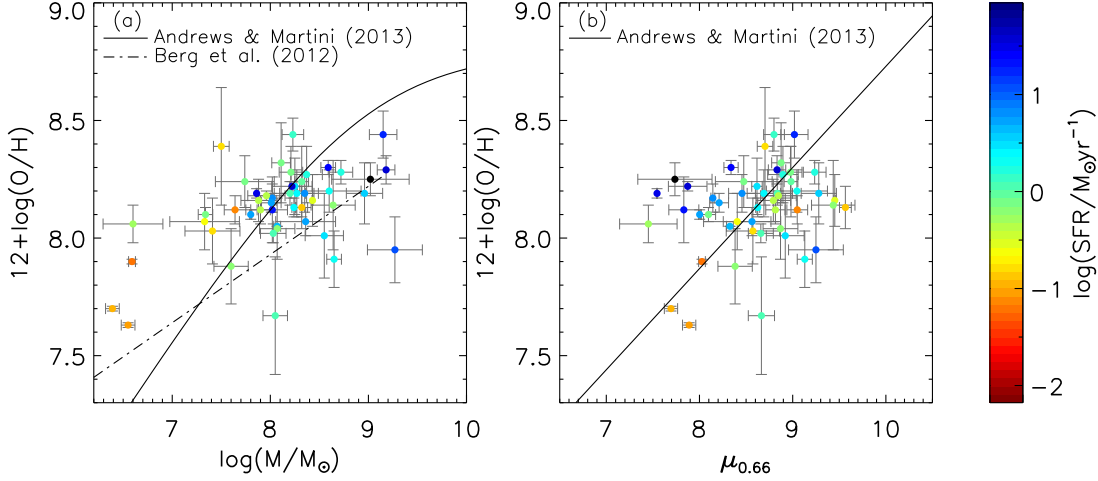}
\caption{The $T_e$ method mass-metallicity relation and fundamental
metallicity relation for our sample galaxies. In both panels, these dot 
symbols of our galaxies are colour-coded by their SFRs. Panel $a$: the solid and
dotted-dashed black lines represent these MZRs
derived from nearby star forming galaxies by \cite{2013ApJ...765..140A} and
\cite{2012ApJ...754...98B}, respectively.
 Panel
$b$: the solid black line represents the FMR relation derived by
\cite{2013ApJ...765..140A}, which assumed the coefficient on $\rm log(SFR)$
is 0.66 with $\rm \mu_{0.66} = log(M) - 0.66 log(SFR)$. The uncertainties 
of stellar masses are presented at 68$\%$ confidence interval limits.} 
 
\label{fmr_rela}
\end{figure}

In panel $a$ of Figure \ref{fmr_rela}, we plot the mass-metallicity relation (MZR)
with  $T_e$-based metallicities for our sample. These dot 
symbols of our galaxies are colour-coded by their SFRs.
For comparison,
we also show the MZRs obtained by \cite{2013ApJ...765..140A} and 
\cite{2012ApJ...754...98B} for their galaxy sample in local universe,
which are shown as solid and dotted-dashed black lines, based on
$T_e$ metallicity calculation.  
 
The MZR in \cite{2012ApJ...754...98B} is a simple linear fit for a small sample 
of low luminosity metal-poor galaxies with stellar masses $\rm log(M/M_{\sun})$
ranging from 5.9 to 9.15. As is shown in panel $a$, the metallicities of 
our 
metal-poor galaxies are systematic higher than the MZR in
\cite{2012ApJ...754...98B} by about 0.25 dex. The MZR of
\cite{2013ApJ...765..140A} is fitted with a asymptotic logarithmic
formula for about two hundred thousands 
nearby star-forming galaxies in stellar mass from $\rm log(M/M_{\sun})=7.4\--10.5$.
Most of our galaxies are in good agreement
with the MZR in \cite{2013ApJ...765..140A},  the 
average and median values of residuals between metallicities and MZR are 
0.0015 dex and 0.025 dex, respectively. 
 We find that the scatter in the MZR from LAMOST 
data, relative to the MZR of \cite{2013ApJ...765..140A}, is 0.28 dex.

\subsection{Mass-metallicity-SFR relation}
\label{fmr}

The mass-metallicity-SFR relation, also  referred to as the 
fundamental metallicity relation (FMR), is proposed by 
\cite{2010MNRAS.408.2115M} to describe the anti-correlation between
metallicity and SFR  at fixed stellar mass. 
\cite{2010MNRAS.408.2115M}
defined a new quantity $\rm \mu_{\alpha} = log(M)-\alpha log(SFR)$ to 
minimize the dispersion in MZR  for 
local galaxies. Using the semi-empirical ``strong-line" metallicity 
calibration of \cite{2008AA...488..463M}, \cite{2010MNRAS.408.2115M} yielded 
$\rm \alpha = 0.32$. However, \cite{2013ApJ...765..140A} found a new value of
$\rm \alpha=0.66$ based on the  $T_e$ metallicity calculation method.
 In this work,
we assume the value of $\rm \alpha=0.66$, since metallicities of our sample galaxy
are also determined with $T_e$ method.

The panel $b$ of Figure \ref{fmr_rela} shows the FMR for our metal-poor 
galaxies. The solid black line represents the FMR relation derived by
\cite{2013ApJ...765..140A}. Most of our galaxies are consistent 
with the FMR,  the average and median values of the 
residuals between metallicities and FMR are 0.002 dex and 0.009 dex, 
respectively. The scatter in the FMR from LAMOST 
data, relative to the FMR of \cite{2013ApJ...765..140A}, is about 0.24 dex.

\section{discussion}
\label{discussion}

\subsection{Comparison with SDSS spectrum}
\label{comp_spec}
\begin{figure}[h]
\center
\includegraphics[width=1.0\textwidth]{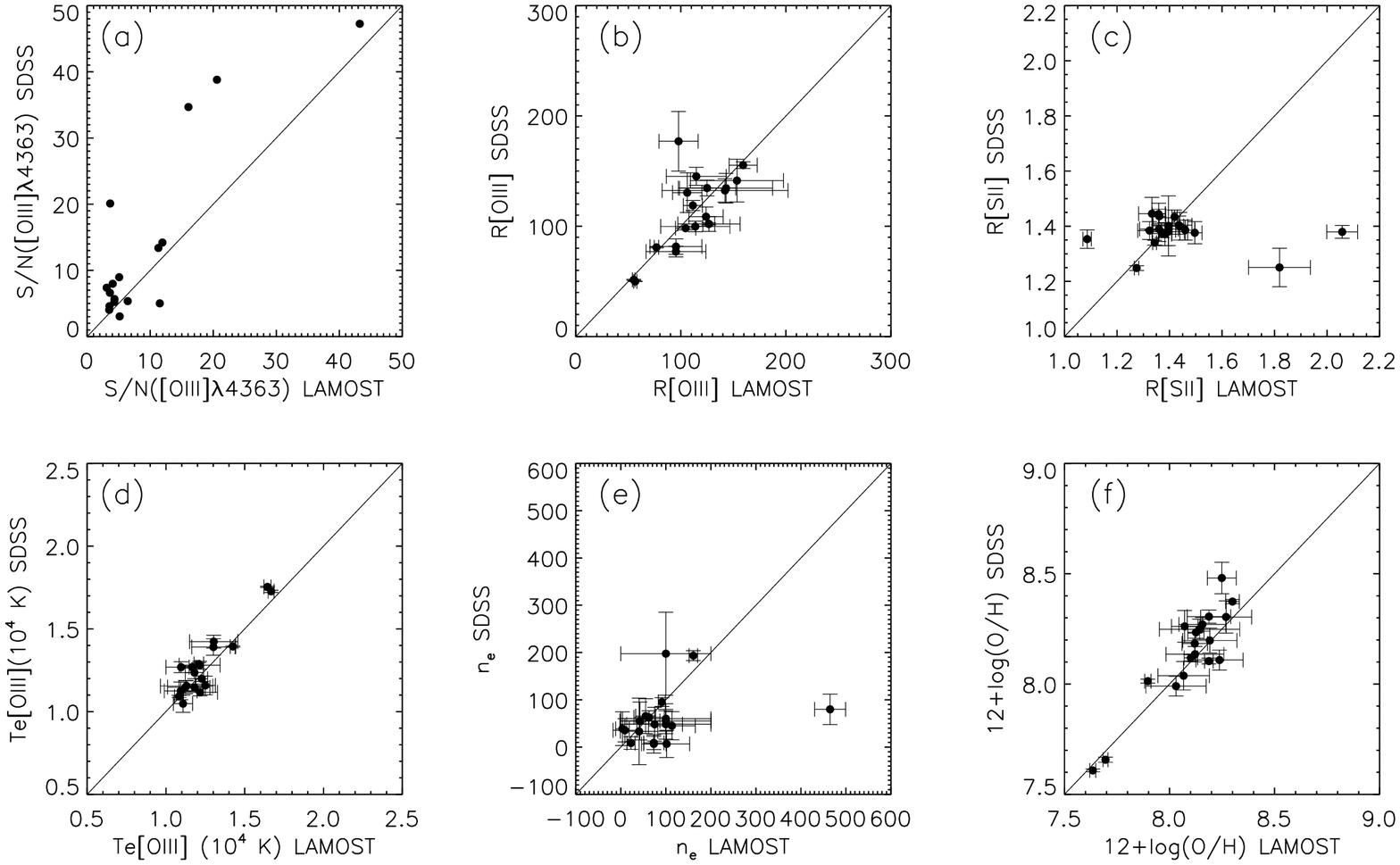}
\caption{The comparisons of signal noise ratios for weak 
\oiii$\lambda4363$ lines, $\oiii$ line fluxes
ratios ($R \oiii = \oiii\lambda\lambda4959,5007/\oiii\lambda4363$), $\sii$ 
line fluxes ratios 
($ R \sii = \sii\lambda6717/\sii\lambda6731$), electron temperatures, 
electron densities and oxygen abundances
derived from LAMOST spectra and from SDSS spectra for 19 galaxies in our sample. 
The solid line indicate equality between the LAMOST and SDSS measurements.}  
\label{lamo_sdss}
\end{figure}

Among our metal-poor galaxy sample, 24 galaxies are also spectrally detected
by SDSS, and are marked with flag "1" in 'SDSS' column of Table 1. We select these 
galaxies from SDSS
DR12 by matching the RA and DEC with our sample  within one arcsec.
  We also obtain the emission line fluxes from these 
24 SDSS galaxy spectra and find that there are 19 spectra with $\oiii\lambda4363$ 
detections above
3$\sigma$. Similarly, we calculate their metallicities with the $T_e$ 
method. Figure \ref{lamo_sdss} 
shows the comparisons of S/Ns for weak $\oiii\lambda4363$ lines, $\oiii$ and $\sii$ line 
fluxes ratios 
($\oiii\lambda\lambda4959,5007/\oiii\lambda4363$, $\sii\lambda6717/\sii\lambda6731$), electron temperatures
($T_e(\oiii)$), electron densities ($n_e$) and oxygen abundances  derived from LAMOST spectra and from
SDSS spectra. 
  The quality of the SDSS 
spectra are generally better than those from
    LAMOST with higher S/N on the weak $\oiii\lambda4363$. Panels $b$ and $d$ 
present strong correlation 
for $\oiii$ ratios and electron temperatures between the LAMOST and SDSS measurements.
Although there are several objects that have large dispersion in the comparison for $\sii$ ratios 
and electron densities, the differences between the final oxygen abundances from these two
measurements are less than 0.01 dex.

\subsection{Comparison with other \oiii$\lambda4363$ galaxy samples}

All of galaxies in our sample  are selected from the local universe 
($0.004 \le z \le 0.14$), and have stellar masses  spanning 
three orders with $\rm 6.39 \le log(M/M_{\sun}) \le 9.27$. The only
XMPG is detected with $\zoh = 7.63 \pm 0.01$ in our sample, however, it has already been found by 
\cite{2012AA...546A.122I}. In the past decades, there have been many efforts to search 
for $\oiii\lambda4363$ galaxies and XMPGs in local universe. For example, 
\cite{2003ApJ...593L..73K} discovered 12 XMPGs with $\rm 7.13 \le \zoh 
\le 7.64$ using SDSS spectroscopy. \cite{2006AA...448..955I} found
6 new XMPGs in 310 $\oiii\lambda4363$ galaxies from SDSS DR3. And
\cite{2012ApJ...754...98B} also researched 19 $\oiii\lambda4363$ 
low luminosity galaxies with MMT telescope.
Additionally, for the intermediate and high redshift universe, 
\cite{2007ApJ...668..853K} mapped 12 XMPGs to $z = 1.0$
with Keck II DEIMOS. \cite{2014ApJ...780..122L} identified 
4 XMPGs in 20 emission-line galaxies with $\oiii\lambda4363$ at 
$z = 0.065-0.90$ by MMT and Keck telescope. 
\cite{2014AA...568L...8A} also discovered 4 XMPGs from 31 low-luminosity
extreme emission line galaxies out to $z = 0.9$ in the VIMOS
Ultra-Deep Survey. Recently, \cite{2015ApJ...805...45L} found 28 
metal-poor galaxies with stellar mass spanning
$\rm 7.1 \times 10^7 - 2.2 \times 10^9 M_{\sun}$ in DEEP2 at redshift $ z \sim 0.8$. 
\cite{2016ApJS..226....5L} also presented a larger sample of 164 galaxies with weak 
$\oiii\lambda4363$ line at $z = 0.1 \-- 1.0$ from the ``Metal Abundances across Cosmic Time'' survey. Compared with these 
samples, the galaxy number in our sample is small, which may be caused by 
 a limiting magnitude selection of LAMOST. However, 
the fraction of galaxies with $\oiii\lambda4363$ detections
in LAMOST data is nearly same as that in SDSS.

\subsection{MZR and FMR}
The MZR relation, which was established originally by \cite{1979AA....80..155L} and developed by
\cite{1987ApJ...317...82G}, \cite{1989ApJ...347..875S}, 
\cite{1991ApJ...379..157B}, \cite{1994ApJ...420...87Z}, 
\cite{2004ApJ...613..898T},
 indicates that the metallicities of galaxies correlate with their stellar masses.
Taking SFR into consideration, \cite{2010MNRAS.408.2115M} found that
metallicity decreases with increasing SFR at low stellar mass, while 
does not depend on SFR at high stellar mass ($\rm log(M/M_{\sun}) \ge
10.7$). However, \cite{2012MNRAS.422..215Y} suggested that high-mass
($\rm log(M/M_{\sun}) \ge 10.4$) galaxies have lower metallicities
when their SFRs are lower. These different results may be caused
by different metallicity calculation methods. In addition, the MZR is also affected by other
physical parameters, such as stellar age and gas fraction. 
\cite{2015MNRAS.446.1449L} found that
the metallicity is strongly dependent on the $D\rm_{n}(4000)$,  which interpreted galaxies with older stellar ages as having
higher metallicities at a fixed stellar mass. \cite{2013AA...550A.115H} found that galaxies with higher
gas fraction have lower metallicities at a fixed mass.
\cite{2010AA...521L..53L} and \cite{2010MNRAS.408.2115M} argued that the MZR 
is in fact a projection
of FMR. In the past years, many efforts (e.g., \cite{2010MNRAS.408.2115M,
 2012ApJ...754...98B, 2013ApJ...765..140A, 2014ApJ...797..126S, 
2015ApJ...805...45L, 2016ApJ...828...67L}) 
have been made to explore the MZR and FMR feasibility from low mass 
to high mass, as well as the evolution with redshift.
In the local universe, the metallicity increases with increasing stellar mass,
and decreases with increasing SFR at a fixed stellar mass when
$\rm log(M/M_{\sun}) \le 10.5$. \cite{2014ApJ...797..126S} found that
the metallicity is anti-corrected with specific SFR regardless of different metallicity
indicators or methods used when $\rm 9.0 \le log(M/M_{\sun}) \le 10.5$, while the dependence is weak or absent for massive galaxies 
when $\rm log(M/M_{\sun}) > 10.5$. \cite{2014ApJ...797..126S} also 
demonstrated that the relative 
specific SFR is a more physically motivated second parameter for the MZR, and found that the overall scatter 
in the FMR relation does not significantly decrease  
relative to the dispersion in the MZR. Recently, 
\cite{2016MNRAS.455.1156B}
reported that the FMR is between stellar mass, metallicity and gas mass
instead of the SFR. In addition, \cite{2016ApJ...823L..24K} measured the 
metallicity of star-forming galaxies based on \cite{2016ApSS.361...61D}
 and \cite{2008AA...488..463M} calibrations, and found that whether the FMR 
exists or not depend on the metallicity measurement method. The dependence on 
metallicity and SFR at high stellar mass is still in argument
\citep{2016ApJ...823L..24K}. 
For the intermediate redshift universe, \cite{2016ApJ...828...67L} showed clearly that the MZR
evolves toward lower metallicity at fixed stellar mass with increasing redshift $z$,
and found a much weaker dependence of MZR on SFR than in the local universe.

In panel $a$ of Figure \ref{fmr_rela}, 
we colour-code our galaxy points with their SFRs.  
Figure \ref{fmr_rela} shows that most of galaxies in our sample have higher metallicities
than that of galaxies in \cite{2012ApJ...754...98B}, but are consistent  with the
result in
\cite{2013ApJ...765..140A}.   The difference  between our 
work and
\cite{2012ApJ...754...98B} may be caused by difference in sample selection and 
calibrations for electron temperatures $T_e(\oiii)$ and $T_e(\oii)$. 
\cite{2013ApJ...765..140A} found that the scatter in MZR for the M-SFR 
stacks with $T_e$-based metallicity is 0.22 dex, while the scatter in the FMR 
is 0.13 dex. The decrease of scatter value in \cite{2013ApJ...765..140A} 
reflects a strong SFR-dependence on the MZR. From 
visual examination, we do not find 
strong dependence of MZR on SFR. However, the scatter in FMR is 0.24 dex, lower 
than the 0.28 dex 
scatter in MZR, suggesting MZR is weakly dependent on SFR. 
We note that the average and median values of metallicity measurement 
uncertainties are 0.09 dex and 0.08 dex, respectively. The average and 
median values on stellar mass measurement uncertainties are about 0.14 dex and 
0.12 dex, respectively.  The larger scatters in 
MZR and FMR compared with \cite{2013ApJ...765..140A} relations may be caused 
by the small galaxy sample size, as well as the measurement uncertainties on
stellar mass.

\section{summary}
\label{summary}  

We inspect all the  92,510 galaxies in LAMOST DR3, DR4 Q1 and Q2, and select  48 galaxies
with $\oiii\lambda4363$  detected at $\ge 3 \sigma$ as our metal-poor galaxy sample. 
 Using the $T_e$ method, we obtain the metallicities 
of these metal-poor galaxies 
with a median $\zoh = 8.16$, spanning from 7.63 to  8.46.    
 The most metal-deficient galaxy in
our sample is ID26 with
$\rm \zoh=7.63 \pm 0.01$, which is the only  XMPG we found, 
but has already been  discovered by \cite{2012AA...546A.122I}. We also confirm
that the galaxy ID33 (RC2 A1228+12) is not an XMPG.

With multiband photometric data from FUV to NIR  and $\ha$ measurements, 
we determine 
the stellar masses and dust-corrected SFRs, based on the SED fitting and 
 reddening corrected $\ha$ luminosities,
respectively. We compare the relationship between stellar mass, 
$T_e$-based metallicity and 
SFR of our galaxies with galaxies in the local universe.
We find that the metallicities of our galaxies are in good agreement with the 
local $T_e$-based MZR in 
\cite{2013ApJ...765..140A}  with average and median values of 
residuals as 0.0015 dex and 0.025 dex, respectively.
However, the MZR in \cite{2012ApJ...754...98B} may be systematic lower
than the metallicities of
our metal-poor galaxies. Assuming the coefficient of $\rm \alpha =0.66$, 
we find most of our galaxies are consistent with the FMR in \cite{2013ApJ...765..140A}.
 However, the scatter in FMR is 0.24 dex, lower than the 0.28 dex 
scatter in MZR, suggesting MZR has a weak dependence on SFR.

\begin{acknowledgements}
We are very grateful to the referee's insightful suggestions 
and comments,  who greatly improve the manuscript 
for this work.
Guoshoujing Telescope (the Large Sky Area Multi-Object Fiber Spectroscopic Telescope LAMOST)
is a National Major Scientific Project built by the Chinese Academy of Sciences. 
Funding for the project has been provided by the National Development and Reform Commission.
LAMOST is operated and managed by the National Astronomical Observatories, 
Chinese Academy of Sciences.

This work is supported by the Strategic Priority Research
Program "The Emergence of Cosmological Structures" of the Chinese Academy
of Sciences (No. XDB09000000), the National Basic Research Program of China
(973 Program)(2015CB857004), and the National Natural Science Foundation of 
China (NSFC, Nos. 11225315, 1320101002, 11433005 and 11421303).
 
\end{acknowledgements}

\bibliography{ms}

\end{document}

%% file: tab1.tex
\begin{center}\tiny \doublerulesep 0.1pt \tabcolsep 1.0pt
\setlength{\extrarowheight}{0.1mm}  
\addtolength{\tabcolsep}{0.1mm}  
\begin{landscape}
\begin{longtable}{@{}lcccrrrccrrrrccrccrr@{}}
\caption[]{The sample of metal-poor galaxies in LAMOST survey} \\
\hline  
\hline ID$\rm ^ a$ & RA$\rm ^ b$ & DEC$\rm ^ b$ & $z\rm ^ b$ &  
\multicolumn{7}{c}{$I(\lambda)/I(\hb)$ $\rm ^ c$} &  $I(\hb) \rm ^ c$ &
$EW(\hb) \rm ^ d$ & $E(B-V) \rm ^ e$  & $ {T_e\oiii \rm ^ f}$ & $ n_e \rm ^ f$ &
$\rm \zoh_{\rm T_e}$ & $\rm log(M)$ &
$\rm log(SFR)$  &  SDSS$\rm ^ g$  \\
\hline
{} & {(deg)} & {(deg)} & {} & $\oii\lambda3727$  & $\oiii\lambda4363$ &
$\oiii\lambda4959$ & $\oiii\lambda5007$   & $\ha$ & $\sii\lambda6717$ &
$\sii\lambda6731$ & {} & {(\AA)} & {(mag)} & ({$10^4$K}) & ($\rm cm^{-3}$)& {} & 
{($M_{\sun}$)}& {($M_{\sun} yr^{-1}$)} & {}  \\  
\endfirsthead
\caption[]{-- continued from previous page}\\ 
\hline
\hline ID$\rm ^ a$ & RA$\rm ^ b$ & DEC$\rm ^ b$ & $z\rm ^ b$ &  
\multicolumn{7}{c}{$I(\lambda)/I(\hb)$ $\rm ^ c$} &  $I(\hb) \rm ^ c$ &
$EW(\hb) \rm ^ d$ & $E(B-V) \rm ^ e$  & $ {T_e\oiii \rm ^ f}$ & $ n_e \rm ^ f$ &
$\rm \zoh_{\rm T_e}$ & $\rm log(M)$ &
$\rm log(SFR)$  &  SDSS$\rm ^ g$  \\
\hline
{} & {(deg)} & {(deg)} & {} & $\oii\lambda3727$  & $\oiii\lambda4363$ &
$\oiii\lambda4959$ & $\oiii\lambda5007$   & $\ha$ & $\sii\lambda6717$ &
$\sii\lambda6731$ & {} & {(\AA)} & {(mag)} & ({$10^4$K}) & ($\rm cm^{-3}$)& {} 
& {($M_{\sun}$)}& {($M_{\sun} yr^{-1}$)} & {}  \\    
\hline  \endhead
\hline \\  \multicolumn{20}{r}{{Continued on next page}} \endfoot
 \endlastfoot
\hline
           1&     0.04352&     4.93125&       0.031&       1.748&       0.049&   
    1.760&       5.432&       2.326&       0.114&       0.063&     1761.90&      
 81.46&        0.00&        1.16&       38.78&        8.22&       8.250&      
-0.556&           0\\
{}&{}&{}&{}&       0.054&       0.013&       0.018&       0.051&       0.022&    
   0.007&       0.007&       15.87&        0.22&        0.01&        0.10&       
42.91&        0.12&       0.050&       0.009&{}\\
           2&     0.22617&    18.50614&       0.055&       2.205&       0.030&   
    1.267&       3.924&       2.135&       0.178&       0.133&     7058.70&      
 46.76&        0.00&        1.05&       99.45&        8.29&       9.180&       
0.527&           0\\
{}&{}&{}&{}&       0.018&       0.004&       0.007&       0.019&       0.010&    
   0.002&       0.001&       33.06&        0.23&        0.01&        0.05&       
15.32&        0.06&       0.090&       0.005&{}\\
           3&    17.64580&     2.11408&       0.016&       2.258&       0.032&   
    1.333&       3.876&       2.685&       0.193&       0.159&     3282.40&      
 34.17&        0.00&        1.07&      234.91&        8.28&       8.720&      
-0.782&           0\\
{}&{}&{}&{}&       0.018&       0.004&       0.004&       0.010&       0.007&    
   0.002&       0.002&        8.34&        0.31&        0.01&        0.05&       
22.81&        0.05&       0.120&       0.003&{}\\
           4&    19.02766&     1.03444&       0.035&       2.164&       0.072&   
    1.335&       3.977&       2.054&       0.130&       0.108&      439.67&      
 43.01&        0.00&        1.47&      272.39&        7.88&       7.600&      
-1.185&           1\\
{}&{}&{}&{}&       0.759&       0.021&       0.022&       0.059&       0.028&    
   0.007&       0.007&        5.88&        0.26&        0.01&        0.19&      
142.65&        0.16&       0.175&       0.014&{}\\
           5&    23.00047&    -2.74083&       0.018&       1.681&       0.056&   
    1.632&       5.063&       2.362&       0.177&       0.152&      797.69&      
 62.08&        0.00&        1.19&      303.92&        8.16&       7.880&      
-1.382&           0\\
{}&{}&{}&{}&       0.062&       0.010&       0.016&       0.043&       0.021&    
   0.008&       0.006&        6.63&        0.48&        0.01&        0.08&      
110.75&        0.09&       0.180&       0.009&{}\\
           6&    31.87386&     4.73166&       0.011&       3.279&       0.019&   
    1.089&       3.269&       2.579&       0.267&       0.193&     5354.90&      
 32.06&        0.00&        0.97&       50.44&        8.44&       8.230&      
-0.863&           0\\
{}&{}&{}&{}&       0.026&       0.003&       0.003&       0.007&       0.006&    
   0.001&       0.001&       11.49&        0.52&        0.01&        0.05&       
 8.88&        0.07&       0.110&       0.002&{}\\
           7&    37.72624&     1.91723&       0.025&       3.333&       0.069&   
    0.997&       2.839&       2.860&       0.379&       0.229&     1143.89&      
 12.62&        0.20&        1.82&       30.48&        7.67&       8.050&      
-0.931&           0\\
{}&{}&{}&{}&       1.790&       0.035&       0.031&       0.068&       0.067&    
   0.017&       0.016&       26.25&        0.58&        0.02&        0.37&       
34.73&        0.25&       0.125&       0.024&{}\\
           8&    39.07656&     1.75273&       0.023&       2.130&       0.053&   
    1.828&       5.206&       2.860&       0.200&       0.144&     6174.59&      
 55.21&        0.25&        1.23&       79.89&        8.17&       8.020&      
-0.187&           0\\
{}&{}&{}&{}&       0.037&       0.009&       0.018&       0.046&       0.026&    
   0.009&       0.008&       54.01&        0.07&        0.01&        0.06&       
65.41&        0.06&       0.015&       0.009&{}\\
           9&    43.15004&    19.68750&       0.029&       3.078&       0.044&   
    1.270&       3.582&       1.836&       0.174&       0.128&     4755.70&      
 27.36&        0.00&        1.22&       73.19&        8.15&       8.010&      
-0.304&           0\\
{}&{}&{}&{}&       0.015&       0.003&       0.003&       0.008&       0.004&    
   0.002&       0.002&       10.31&        0.16&        0.01&        0.04&       
21.25&        0.04&       0.100&       0.002&{}\\
          10&    47.66477&     5.23287&       0.064&       2.962&       0.015&   
    1.111&       3.332&       2.853&       0.242&       0.182&     3494.70&      
 35.56&        0.00&        0.96&      113.42&        8.44&       9.150&       
0.199&           0\\
{}&{}&{}&{}&       0.053&       0.005&       0.011&       0.029&       0.025&    
   0.002&       0.002&       30.42&        0.23&        0.01&        0.07&       
15.29&        0.10&       0.140&       0.009&{}\\
          11&    47.91879&     2.57282&       0.020&       1.685&       0.098&   
    2.115&       6.153&       2.582&       0.139&       0.086&     6170.00&      
103.02&        0.00&        1.37&       15.47&        8.05&       8.070&      
-0.389&           0\\
{}&{}&{}&{}&       0.021&       0.003&       0.005&       0.013&       0.005&    
   0.001&       0.001&       12.86&        6.06&        0.01&        0.00&       
 4.26&        0.00&       0.035&       0.002&{}\\
          12&    49.35046&     3.50387&       0.039&       2.179&       0.047&   
    1.523&       4.753&       2.860&       0.177&       0.129&     2297.92&      
 81.85&        0.09&        1.17&       67.05&        8.19&       8.350&      
-0.161&           0\\
{}&{}&{}&{}&       0.181&       0.009&       0.010&       0.027&       0.016&    
   0.005&       0.004&       12.85&        0.73&        0.01&        0.08&       
44.49&        0.08&       0.140&       0.006&{}\\
          13&    51.95898&     1.02635&       0.109&       2.973&       0.030&   
    1.111&       3.332&       2.816&       0.443&       0.300&     1148.30&      
 45.87&        0.00&        1.11&        6.72&        8.25&       9.020&       
1.945&           1\\
{}&{}&{}&{}&       1.280&       0.232&       0.193&       0.458&       0.403&    
   0.154&       0.132&      151.78&        0.21&        0.01&        0.06&       
 7.13&        0.07&       0.395&       0.005&{}\\
          14&   123.48118&    23.14714&       0.015&       3.171&       0.052&   
    1.245&       3.601&       2.339&       0.226&       0.161&     1515.70&      
 50.33&        0.00&        1.33&       45.20&        8.06&       6.600&      
-1.290&           0\\
{}&{}&{}&{}&       0.077&       0.008&       0.007&       0.017&       0.011&    
   0.003&       0.003&        6.85&        0.31&        0.01&        0.09&       
29.09&        0.08&       0.305&       0.005&{}\\
          15&   129.35571&    37.51241&       0.042&       2.942&       0.050&   
    1.118&       3.206&       2.639&       0.340&       0.230&      942.03&      
 30.75&        0.00&        1.34&       25.58&        8.01&       8.550&      
-0.561&           1\\
{}&{}&{}&{}&       0.178&       0.018&       0.017&       0.038&       0.032&    
   0.009&       0.009&       10.81&        0.22&        0.01&        0.21&       
26.69&        0.18&       0.195&       0.012&{}\\
          16&   136.71599&    41.36413&       0.135&       3.205&       0.056&   
    1.107&       3.224&       2.166&       0.231&       0.182&      446.71&      
 32.70&        0.00&        1.43&      187.71&        7.95&       9.270&       
0.031&           1\\
{}&{}&{}&{}&       0.077&       0.016&       0.030&       0.070&       0.045&    
   0.012&       0.008&        9.18&        0.42&        0.02&        0.17&      
100.22&        0.14&       0.280&       0.021&{}\\
          17&   139.46461&    40.97089&       0.024&       2.608&       0.031&   
    1.403&       3.912&       2.207&       0.262&       0.189&      721.63&      
 42.77&        0.00&        1.06&       52.69&        8.32&       8.110&      
-1.160&           0\\
{}&{}&{}&{}&       0.075&       0.013&       0.015&       0.035&       0.020&    
   0.004&       0.004&        6.30&        0.10&        0.01&        0.14&       
23.70&        0.17&       0.095&       0.009&{}\\
          18&   139.95418&     4.48470&       0.012&       2.452&       0.057&   
    1.376&       4.164&       2.338&       0.227&       0.167&     2868.10&      
 34.57&        0.00&        1.32&       73.19&        8.04&       8.070&      
-1.211&           0\\
{}&{}&{}&{}&       0.194&       0.015&       0.016&       0.041&       0.023&    
   0.005&       0.004&       27.82&        0.09&        0.01&        0.14&       
38.93&        0.13&       0.180&       0.010&{}\\
          19&   140.17699&     5.73677&       0.038&       1.029&       0.082&   
    2.450&       7.272&       2.860&       0.100&       0.071&    11516.86&      
 96.12&        0.23&        1.20&       38.78&        8.22&       8.220&       
0.525&           0\\
{}&{}&{}&{}&       0.014&       0.003&       0.007&       0.019&       0.008&    
   0.001&       0.001&       30.64&        0.31&        0.01&        0.02&       
23.27&        0.02&       0.205&       0.003&{}\\
          20&   140.92796&     3.36689&       0.012&       2.760&       0.050&   
    1.340&       3.788&       2.312&       0.210&       0.150&     1090.50&      
 34.17&        0.00&        1.30&       52.69&        8.07&       7.330&      
-1.629&           1\\
{}&{}&{}&{}&       0.086&       0.014&       0.015&       0.035&       0.021&    
   0.006&       0.005&        9.79&        0.16&        0.01&        0.14&       
41.58&        0.12&       0.355&       0.010&{}\\
          21&   142.63454&    34.43079&       0.017&       1.689&       0.045&   
    1.443&       4.180&       2.860&       0.190&       0.104&     1297.69&      
 34.84&        0.01&        1.10&       56.29&        8.24&       7.740&      
-1.114&           1\\
{}&{}&{}&{}&       0.037&       0.010&       0.012&       0.029&       0.020&    
   0.006&       0.006&        8.78&        0.21&        0.01&        0.10&       
66.58&        0.11&       0.300&       0.007&{}\\
          22&   145.71992&    35.79055&       0.015&       1.676&       0.055&   
    1.538&       4.217&       2.210&       0.166&       0.081&     2180.30&      
100.80&        0.00&        1.21&       19.54&        8.10&       7.339&      
-1.147&           1\\
{}&{}&{}&{}&       0.021&       0.005&       0.006&       0.015&       0.008&    
   0.002&       0.002&        7.31&        0.15&        0.01&        0.03&       
10.87&        0.03&       0.074&       0.004&{}\\
          23&   146.00691&    50.87919&       0.038&       2.683&       0.028&   
    1.048&       2.923&       2.471&       0.320&       0.235&     1322.90&      
 37.23&        0.00&        1.12&       73.19&        8.19&       8.960&      
-0.485&           1\\
{}&{}&{}&{}&       0.036&       0.009&       0.008&       0.018&       0.015&    
   0.004&       0.003&        7.63&        0.64&        0.01&        0.11&       
22.38&        0.13&       0.185&       0.006&{}\\
          24&   147.03999&     2.52892&       0.021&       2.681&       0.032&   
    1.348&       4.014&       2.844&       0.279&       0.194&     2646.10&      
 35.59&        0.00&        1.18&       30.48&        8.19&       8.270&      
-0.638&           1\\
{}&{}&{}&{}&       0.086&       0.009&       0.010&       0.026&       0.019&    
   0.005&       0.006&       16.94&        0.09&        0.01&        0.10&       
26.41&        0.10&       0.175&       0.007&{}\\
          25&   148.70041&    38.45017&       0.017&       3.040&       0.024&   
    0.939&       2.683&       2.440&       0.370&       0.269&     1971.30&      
 29.46&        0.00&        1.09&       62.79&        8.24&       8.320&      
-1.003&           1\\
{}&{}&{}&{}&       0.036&       0.008&       0.008&       0.016&       0.014&    
   0.004&       0.004&       11.02&        0.17&        0.01&        0.13&       
17.34&        0.15&       0.070&       0.006&{}\\
          26&   154.10216&    37.91277&       0.004&       0.687&       0.097&   
    1.346&       4.106&       2.860&       0.074&       0.056&     3778.46&      
138.04&        0.01&        1.64&      115.85&        7.63&       6.550&      
-2.029&           1\\
{}&{}&{}&{}&       0.022&       0.003&       0.003&       0.007&       0.005&    
   0.001&       0.002&        6.15&        0.19&        0.01&        0.02&       
52.03&        0.01&       0.070&       0.002&{}\\
          27&   157.45534&    16.18091&       0.011&       3.230&       0.040&   
    1.187&       3.468&       2.007&       0.278&       0.203&     1701.80&      
 27.67&        0.00&        1.22&       64.22&        8.16&       8.430&      
-1.548&           1\\
{}&{}&{}&{}&       0.059&       0.010&       0.012&       0.028&       0.016&    
   0.005&       0.005&       13.14&        0.21&        0.01&        0.11&       
33.81&        0.11&       0.005&       0.008&{}\\
          28&   161.47824&     1.06829&       0.026&       2.243&       0.045&   
    1.549&       4.167&       2.860&       0.216&       0.161&    22426.28&      
108.84&        0.18&        1.17&       91.11&        8.19&       7.860&       
0.478&           1\\
{}&{}&{}&{}&       0.012&       0.002&       0.003&       0.006&       0.004&    
   0.001&       0.001&       33.64&        0.24&        0.01&        0.02&       
 7.99&        0.02&       0.005&       0.001&{}\\
          29&   162.63853&    22.31531&       0.046&       2.084&       0.052&   
    1.394&       3.964&       2.826&       0.313&       0.218&     1269.00&      
 39.66&        0.00&        1.25&       16.87&        8.07&       8.360&      
-0.308&           1\\
{}&{}&{}&{}&       0.030&       0.007&       0.008&       0.019&       0.014&    
   0.004&       0.004&        6.06&        0.19&        0.01&        0.06&       
15.18&        0.06&       0.305&       0.005&{}\\
          30&   163.09535&    32.63737&       0.005&       2.665&       0.028&   
    1.020&       2.611&       2.412&       0.319&       0.225&     1828.20&      
 20.58&        0.00&        1.18&       22.43&        8.12&       7.640&      
-2.135&           1\\
{}&{}&{}&{}&       0.022&       0.004&       0.004&       0.007&       0.006&    
   0.002&       0.001&        4.74&        0.06&        0.01&        0.06&       
 8.89&        0.06&       0.110&       0.002&{}\\
          31&   170.34760&     6.66713&       0.009&       2.816&       0.064&   
    1.052&       3.069&       2.860&       0.257&       0.193&     1169.43&      
 27.91&        0.01&        1.30&       99.45&        8.03&       7.410&      
-1.762&           1\\
{}&{}&{}&{}&       0.073&       0.013&       0.012&       0.027&       0.025&    
   0.006&       0.006&        9.96&        0.05&        0.01&        0.15&       
51.12&        0.14&       0.280&       0.009&{}\\
          32&   186.62852&    28.84698&       0.027&       3.039&       0.046&   
    0.929&       2.844&       2.860&       0.423&       0.315&     1498.23&      
 25.58&        0.03&        1.44&       89.21&        7.91&       8.650&      
-0.727&           0\\
{}&{}&{}&{}&       0.062&       0.012&       0.011&       0.025&       0.024&    
   0.007&       0.006&       12.52&        0.35&        0.01&        0.15&       
32.31&        0.12&       0.075&       0.009&{}\\
          33&   187.70251&    12.04523&       0.004&       1.118&       0.109&   
    1.565&       4.442&       2.697&       0.109&       0.080&     4437.80&      
 85.03&        0.00&        1.67&       73.19&        7.70&       6.390&      
-1.976&           1\\
{}&{}&{}&{}&       0.016&       0.002&       0.003&       0.008&       0.005&    
   0.001&       0.001&        7.57&        0.16&        0.01&        0.02&       
26.96&        0.01&       0.070&       0.002&{}\\
          34&   188.68443&    10.71915&       0.032&       2.124&       0.052&   
    1.531&       4.283&       2.740&       0.185&       0.171&     1392.00&      
 67.30&        0.00&        1.23&      464.63&        8.13&       8.240&      
-0.583&           1\\
{}&{}&{}&{}&       0.017&       0.004&       0.005&       0.012&       0.008&    
   0.002&       0.002&        3.93&        0.43&        0.01&        0.04&       
34.42&        0.04&       0.020&       0.003&{}\\
          35&   189.23994&    10.13030&       0.027&       1.485&       0.039&   
    1.366&       4.083&       2.856&       0.249&       0.183&      445.10&      
 37.11&        0.00&        1.13&       85.32&        8.14&       8.640&      
-1.205&           1\\
{}&{}&{}&{}&       0.064&       0.016&       0.019&       0.049&       0.034&    
   0.008&       0.008&        5.16&        0.31&        0.01&        0.17&       
62.00&        0.19&       0.225&       0.012&{}\\
          36&   189.52870&    10.16557&       0.004&       0.642&       0.099&   
    1.953&       5.567&       2.703&       0.089&       0.062&     2579.20&      
156.19&        0.00&        1.43&       16.87&        7.90&       6.590&      
-2.177&           1\\
{}&{}&{}&{}&       0.019&       0.003&       0.004&       0.011&       0.006&    
   0.001&       0.001&        5.14&        0.05&        0.01&        0.02&       
15.98&        0.01&       0.035&       0.002&{}\\
          37&   189.73721&    38.09029&       0.007&       2.820&       0.032&   
    1.046&       2.960&       2.729&       0.367&       0.257&     1454.40&      
 11.09&        0.00&        1.19&       20.55&        8.13&       8.320&      
-1.887&           1\\
{}&{}&{}&{}&       0.040&       0.007&       0.007&       0.014&       0.013&    
   0.004&       0.004&        6.43&        0.23&        0.01&        0.09&       
15.20&        0.09&       0.060&       0.005&{}\\
          38&   196.86958&    54.44713&       0.033&       2.544&       0.035&   
    1.524&       4.133&       2.860&       0.242&       0.190&    10397.45&      
 73.00&        0.03&        1.08&      162.86&        8.30&       8.590&       
0.381&           1\\
{}&{}&{}&{}&       0.014&       0.003&       0.004&       0.009&       0.006&    
   0.001&       0.001&       22.36&        0.13&        0.01&        0.03&       
 9.56&        0.03&       0.075&       0.002&{}\\
          39&   198.22404&    17.20867&       0.052&       2.129&       0.049&   
    1.436&       4.157&       2.860&       0.210&       0.150&     3750.41&      
 70.01&        0.21&        1.21&       57.52&        8.12&       8.020&       
0.285&           1\\
{}&{}&{}&{}&       0.071&       0.014&       0.016&       0.039&       0.027&    
   0.005&       0.006&       34.19&        0.11&        0.01&        0.13&       
44.45&        0.14&       0.155&       0.009&{}\\
          40&   212.67758&    38.71842&       0.025&       3.061&       0.028&   
    1.114&       3.227&       2.860&       0.378&       0.274&     1216.35&      
 23.52&        0.00&        1.09&       52.69&        8.27&       8.370&      
-0.796&           1\\
{}&{}&{}&{}&       0.041&       0.008&       0.008&       0.019&       0.017&    
   0.005&       0.004&        6.84&        0.26&        0.01&        0.10&       
22.31&        0.12&       0.160&       0.006&{}\\
          41&   220.85265&    28.30123&       0.013&       2.490&       0.029&   
    1.199&       3.575&       2.551&       0.291&       0.206&     2186.30&      
 34.68&        0.00&        1.07&       27.92&        8.28&       8.210&      
-1.161&           0\\
{}&{}&{}&{}&       0.038&       0.007&       0.007&       0.018&       0.013&    
   0.003&       0.003&       10.51&        0.67&        0.01&        0.09&       
16.76&        0.11&       0.120&       0.005&{}\\
          42&   222.59503&    11.40265&       0.006&       1.260&       0.036&   
    1.543&       4.234&       1.543&       0.115&       0.214&     2445.60&      
 68.75&        0.00&        1.10&      475.04&        8.39&       7.500&      
-1.823&           0\\
{}&{}&{}&{}&      12.190&       0.007&       0.008&       0.019&       0.007&    
   0.003&       0.002&       10.98&        0.37&        0.01&        0.07&       
48.24&        0.25&       0.080&       0.005&{}\\
          43&   321.52011&     8.68640&       0.010&       1.850&       0.061&   
    1.617&       4.840&       2.809&       0.200&       0.144&     2200.90&      
 49.17&        0.00&        1.24&       44.22&        8.12&       7.900&      
-1.386&           0\\
{}&{}&{}&{}&       0.030&       0.007&       0.009&       0.023&       0.014&    
   0.002&       0.002&       10.48&        0.21&        0.01&        0.06&       
22.79&        0.06&       0.100&       0.005&{}\\
          44&   325.30099&     8.68733&       0.032&       3.758&       0.033&   
    1.055&       3.104&       2.678&       0.343&       0.248&     1105.70&      
 23.31&        0.00&        1.19&       55.05&        8.20&       8.600&      
-0.675&           0\\
{}&{}&{}&{}&       0.032&       0.007&       0.006&       0.016&       0.012&    
   0.003&       0.003&        4.89&        0.34&        0.01&        0.09&       
15.64&        0.09&       0.325&       0.004&{}\\
          45&   327.59015&     7.60371&       0.027&       1.501&       0.094&   
    1.929&       5.831&       2.420&       0.109&       0.082&     1003.90&      
 75.16&        0.00&        1.37&      116.11&        8.02&       8.030&      
-0.952&           0\\
{}&{}&{}&{}&       0.098&       0.009&       0.012&       0.033&       0.014&    
   0.003&       0.004&        5.56&        0.24&        0.01&        0.05&       
72.01&        0.04&       0.130&       0.006&{}\\
          46&   336.10513&     6.25363&       0.016&       2.475&       0.054&   
    1.631&       4.790&       2.327&       0.156&       0.130&     2647.40&      
 63.85&        0.00&        1.20&      260.72&        8.19&       8.200&      
-0.962&           0\\
{}&{}&{}&{}&       0.010&       0.007&       0.008&       0.020&       0.010&    
   0.003&       0.003&       10.75&        0.34&        0.01&        0.05&       
53.29&        0.05&       0.160&       0.004&{}\\
          47&   341.07831&     5.36256&       0.006&       2.681&       0.046&   
    1.436&       4.053&       2.669&       0.228&       0.177&     6841.20&      
 55.27&        0.00&        1.19&      154.47&        8.18&       7.960&      
-1.344&           0\\
{}&{}&{}&{}&       0.022&       0.003&       0.003&       0.007&       0.005&    
   0.001&       0.001&       11.57&        0.46&        0.01&        0.02&       
10.14&        0.02&       0.190&       0.002&{}\\
          48&   352.58768&     5.52682&       0.014&       2.313&       0.057&   
    1.646&       4.242&       2.860&       0.234&       0.173&    13399.27&      
 66.73&        0.10&        1.26&       79.89&        8.10&       7.800&      
-0.306&           0\\
{}&{}&{}&{}&       0.018&       0.003&       0.003&       0.007&       0.005&    
   0.001&       0.002&       21.79&        0.29&        0.01&        0.02&       
14.80&        0.02&       0.010&       0.002&{}\\

\hline
\label{table2}
\end{longtable}
\begin{flushleft}
{\sc Notes:} --- Basic information, emission line fluxes, electron temperatures and oxygen abundances 
for our sample galaxies. For every object, the first (second) line presents the parameter (error) values.\\
$\rm ^a$ 'ID' is the serial number for every object and it will be referred to throughout this paper. \\
$\rm ^b$  The right ascension (J2000) and declination (J2000) of our sample 
galaxies are given in units of degrees.
The RA, DEC, and redshift are obtained from the header of the spectral FITS 
files.\\ 
$\rm ^c$  Reddening corrected emission line fluxes for our sample galaxies measured from the LAMOST
spectra are relative to $\hb$. The $\hb$ fluxes are reported in units of $\rm 10^{-17}erg \ s^{-1}cm^{-2}$.\\
$\rm ^d$  The $\hb$ equivalent widths are given in units of $\rm \AA$, assuming the mean values of 
observed flux intensities within 50$\rm \AA$ wide component around the $\hb$ as the
continuum spectral flux intensities. \\
$\rm ^e$ The nebular color excesses are derived from the observed flux ratios $\ha/\hb$,
and are assumed to be zero when the observed flux ratios $\ha/\hb$ are less than 2.86. \\
$\rm ^f$ Electron temperatures are computed from the oxygen emission line ratios
$\oiii\lambda\lambda 4959,5007/\oiii\lambda4363$. Electron densities are calculated
from an iterative process with $\oiii\lambda\lambda 4959,5007/\oiii\lambda4363$ and
$\sii\lambda6717/\sii\lambda6731$ ratios. \\
$\rm ^g$ The flag numbers indicate the spectral detected states for our objects
with SDSS. "1" ("0") represent this object has (not) been spectroscopically 
detected by 
SDSS.

\end{flushleft}
\end{landscape}
\end{center}

%% file: ms.bbl
\begin{thebibliography}{74}
\providecommand{\natexlab}[1]{#1}
\providecommand{\selectlanguage}[1]{\relax}

\bibitem[{{Abazajian} et~al.(2004){Abazajian}, {Adelman-McCarthy},
  {Ag{\"u}eros} et~al.}]{2004AJ....128..502A}
{Abazajian}, K., {Adelman-McCarthy}, J.~K., {Ag{\"u}eros}, M.~A., et~al. 2004,
  \aj, 128, 502

\bibitem[{{Alam} et~al.(2015){Alam}, {Albareti}, {Allende Prieto}
  et~al.}]{2015ApJS..219...12A}
{Alam}, S., {Albareti}, F.~D., {Allende Prieto}, C., et~al. 2015, \apjs, 219,
  12

\bibitem[{{Allende Prieto} et~al.(2001){Allende Prieto}, {Lambert}, \&
  {Asplund}}]{2001ApJ...556L..63A}
{Allende Prieto}, C., {Lambert}, D.~L., \& {Asplund}, M. 2001, \apjl, 556, L63

\bibitem[{{Aller}(1984)}]{1984ASSL..112.....A}
{Aller}, L.~H., ed. 1984, {Physics of thermal gaseous nebulae},
  \emph{Astrophysics and Space Science Library}, vol. 112

\bibitem[{{Amor{\'{\i}}n} et~al.(2014){Amor{\'{\i}}n}, {Sommariva},
  {Castellano} et~al.}]{2014AA...568L...8A}
{Amor{\'{\i}}n}, R., {Sommariva}, V., {Castellano}, M., et~al. 2014, \aap, 568,
  L8

\bibitem[{{Andrews} \& {Martini}(2013)}]{2013ApJ...765..140A}
{Andrews}, B.~H., \& {Martini}, P. 2013, \apj, 765, 140

\bibitem[{{Baldwin} et~al.(1981){Baldwin}, {Phillips}, \&
  {Terlevich}}]{1981PASP...93..817B}
{Baldwin}, A., {Phillips}, M.~M., \& {Terlevich}, R. 1981, \pasp, 93, 817

\bibitem[{{Berg} et~al.(2012){Berg}, {Skillman}, {Marble}
  et~al.}]{2012ApJ...754...98B}
{Berg}, D.~A., {Skillman}, E.~D., {Marble}, A.~R., et~al. 2012, \apj, 754, 98

\bibitem[{{Bianchi} et~al.(2014){Bianchi}, {Conti}, \&
  {Shiao}}]{2014AdSpR..53..900B}
{Bianchi}, L., {Conti}, A., \& {Shiao}, B. 2014, Advances in Space Research,
  53, 900

\bibitem[{{Bothwell} et~al.(2016){Bothwell}, {Maiolino}, {Peng}
  et~al.}]{2016MNRAS.455.1156B}
{Bothwell}, M.~S., {Maiolino}, R., {Peng}, Y., et~al. 2016, \mnras, 455, 1156

\bibitem[{{Brinchmann} et~al.(2004){Brinchmann}, {Charlot}, {White}
  et~al.}]{2004MNRAS.351.1151B}
{Brinchmann}, J., {Charlot}, S., {White}, S.~D.~M., et~al. 2004, \mnras, 351,
  1151

\bibitem[{{Brodie} \& {Huchra}(1991)}]{1991ApJ...379..157B}
{Brodie}, J.~P., \& {Huchra}, J.~P. 1991, \apj, 379, 157

\bibitem[{{Brorby} et~al.(2014){Brorby}, {Kaaret}, \&
  {Prestwich}}]{2014MNRAS.441.2346B}
{Brorby}, M., {Kaaret}, P., \& {Prestwich}, A. 2014, \mnras, 441, 2346

\bibitem[{{Bruzual} \& {Charlot}(2003)}]{2003MNRAS.344.1000B}
{Bruzual}, G., \& {Charlot}, S. 2003, \mnras, 344, 1000

\bibitem[{{Calzetti} et~al.(2000){Calzetti}, {Armus}, {Bohlin}
  et~al.}]{2000ApJ...533..682C}
{Calzetti}, D., {Armus}, L., {Bohlin}, R.~C., et~al. 2000, \apj, 533, 682

\bibitem[{{Chabrier}(2003)}]{2003PASP..115..763C}
{Chabrier}, G. 2003, \pasp, 115, 763

\bibitem[{{Cid Fernandes} et~al.(2005){Cid Fernandes}, {Mateus}, {Sodr{\'e}},
  {Stasi{\'n}ska}, \& {Gomes}}]{2005MNRAS.358..363C}
{Cid Fernandes}, R., {Mateus}, A., {Sodr{\'e}}, L., {Stasi{\'n}ska}, G., \&
  {Gomes}, J.~M. 2005, \mnras, 358, 363

\bibitem[{{Cui} et~al.(2012){Cui}, {Zhao}, {Chu} et~al.}]{2012RAA....12.1197C}
{Cui}, X.-Q., {Zhao}, Y.-H., {Chu}, Y.-Q., et~al. 2012, Research in Astronomy
  and Astrophysics, 12, 1197

\bibitem[{{Dopita} et~al.(2016){Dopita}, {Kewley}, {Sutherland}, \&
  {Nicholls}}]{2016ApSS.361...61D}
{Dopita}, M.~A., {Kewley}, L.~J., {Sutherland}, R.~S., \& {Nicholls}, D.~C.
  2016, \apss, 361, 61

\bibitem[{{Doyle} et~al.(2005){Doyle}, {Drinkwater}, {Rohde}
  et~al.}]{2005MNRAS.361...34D}
{Doyle}, M.~T., {Drinkwater}, M.~J., {Rohde}, D.~J., et~al. 2005, \mnras, 361,
  34

\bibitem[{{Froese Fischer} \& {Tachiev}(2004)}]{2004ADNDT..87....1F}
{Froese Fischer}, C., \& {Tachiev}, G. 2004, Atomic Data and Nuclear Data
  Tables, 87, 1

\bibitem[{{Garnett}(1992)}]{1992AJ....103.1330G}
{Garnett}, D.~R. 1992, \aj, 103, 1330

\bibitem[{{Garnett} \& {Shields}(1987)}]{1987ApJ...317...82G}
{Garnett}, D.~R., \& {Shields}, G.~A. 1987, \apj, 317, 82

\bibitem[{Hu et~al.(2016)Hu, Su, \& Kong}]{1674-4527-16-3-006}
Hu, N., Su, S.-S., \& Kong, X. 2016, Research in Astronomy and Astrophysics,
  16, 006

\bibitem[{{Hughes} et~al.(2013){Hughes}, {Cortese}, {Boselli}, {Gavazzi}, \&
  {Davies}}]{2013AA...550A.115H}
{Hughes}, T.~M., {Cortese}, L., {Boselli}, A., {Gavazzi}, G., \& {Davies},
  J.~I. 2013, \aap, 550, A115

\bibitem[{{Hummer} \& {Storey}(1987)}]{1987MNRAS.224..801H}
{Hummer}, D.~G., \& {Storey}, P.~J. 1987, \mnras, 224, 801

\bibitem[{{Izotov} et~al.(2006){Izotov}, {Stasi{\'n}ska}, {Meynet}, {Guseva},
  \& {Thuan}}]{2006AA...448..955I}
{Izotov}, Y.~I., {Stasi{\'n}ska}, G., {Meynet}, G., {Guseva}, N.~G., \&
  {Thuan}, T.~X. 2006, \aap, 448, 955

\bibitem[{{Izotov} \& {Thuan}(2004)}]{2004ApJ...616..768I}
{Izotov}, Y.~I., \& {Thuan}, T.~X. 2004, \apj, 616, 768

\bibitem[{{Izotov} et~al.(2012{\natexlab{a}}){Izotov}, {Thuan}, \&
  {Guseva}}]{2012AA...546A.122I}
{Izotov}, Y.~I., {Thuan}, T.~X., \& {Guseva}, N.~G. 2012{\natexlab{a}}, \aap,
  546, A122

\bibitem[{{Izotov} et~al.(2012{\natexlab{b}}){Izotov}, {Thuan}, \&
  {Privon}}]{2012MNRAS.427.1229I}
{Izotov}, Y.~I., {Thuan}, T.~X., \& {Privon}, G. 2012{\natexlab{b}}, \mnras,
  427, 1229

\bibitem[{{Kakazu} et~al.(2007){Kakazu}, {Cowie}, \&
  {Hu}}]{2007ApJ...668..853K}
{Kakazu}, Y., {Cowie}, L.~L., \& {Hu}, E.~M. 2007, \apj, 668, 853

\bibitem[{{Kashino} et~al.(2016){Kashino}, {Renzini}, {Silverman}, \&
  {Daddi}}]{2016ApJ...823L..24K}
{Kashino}, D., {Renzini}, A., {Silverman}, J.~D., \& {Daddi}, E. 2016, \apjl,
  823, L24

\bibitem[{{Kauffmann} et~al.(2003){Kauffmann}, {Heckman}, {White}
  et~al.}]{2003MNRAS.341...33K}
{Kauffmann}, G., {Heckman}, T.~M., {White}, S.~D.~M., et~al. 2003, \mnras, 341,
  33

\bibitem[{{Kewley} et~al.(2001){Kewley}, {Dopita}, {Sutherland}, {Heisler}, \&
  {Trevena}}]{2001ApJ...556..121K}
{Kewley}, L.~J., {Dopita}, M.~A., {Sutherland}, R.~S., {Heisler}, C.~A., \&
  {Trevena}, J. 2001, \apj, 556, 121

\bibitem[{{Kisielius} et~al.(2009){Kisielius}, {Storey}, {Ferland}, \&
  {Keenan}}]{2009MNRAS.397..903K}
{Kisielius}, R., {Storey}, P.~J., {Ferland}, G.~J., \& {Keenan}, F.~P. 2009,
  \mnras, 397, 903

\bibitem[{{Kniazev} et~al.(2003){Kniazev}, {Grebel}, {Hao}
  et~al.}]{2003ApJ...593L..73K}
{Kniazev}, A.~Y., {Grebel}, E.~K., {Hao}, L., et~al. 2003, \apjl, 593, L73

\bibitem[{{Kriek} et~al.(2009){Kriek}, {van Dokkum}, {Labb{\'e}}
  et~al.}]{2009ApJ...700..221K}
{Kriek}, M., {van Dokkum}, P.~G., {Labb{\'e}}, I., et~al. 2009, \apj, 700, 221

\bibitem[{{Kunth} \& {{\"O}stlin}(2000)}]{2000AARv..10....1K}
{Kunth}, D., \& {{\"O}stlin}, G. 2000, \aapr, 10, 1

\bibitem[{{Lara-L{\'o}pez} et~al.(2010){Lara-L{\'o}pez}, {Cepa}, {Bongiovanni}
  et~al.}]{2010AA...521L..53L}
{Lara-L{\'o}pez}, M.~A., {Cepa}, J., {Bongiovanni}, A., et~al. 2010, \aap, 521,
  L53

\bibitem[{{Lequeux} et~al.(1979){Lequeux}, {Peimbert}, {Rayo}, {Serrano}, \&
  {Torres-Peimbert}}]{1979AA....80..155L}
{Lequeux}, J., {Peimbert}, M., {Rayo}, J.~F., {Serrano}, A., \&
  {Torres-Peimbert}, S. 1979, \aap, 80, 155

\bibitem[{{Lian} et~al.(2016){Lian}, {Hu}, {Fang}, {Ye}, \&
  {Kong}}]{2016ApJ...819...73L}
{Lian}, J., {Hu}, N., {Fang}, G., {Ye}, C., \& {Kong}, X. 2016, \apj, 819, 73

\bibitem[{{Lian} et~al.(2015){Lian}, {Li}, {Yan}, \&
  {Kong}}]{2015MNRAS.446.1449L}
{Lian}, J.~H., {Li}, J.~R., {Yan}, W., \& {Kong}, X. 2015, \mnras, 446, 1449

\bibitem[{{L{\'o}pez-S{\'a}nchez} et~al.(2012){L{\'o}pez-S{\'a}nchez},
  {Dopita}, {Kewley} et~al.}]{2012MNRAS.426.2630L}
{L{\'o}pez-S{\'a}nchez}, {\'A}.~R., {Dopita}, M.~A., {Kewley}, L.~J., et~al.
  2012, \mnras, 426, 2630

\bibitem[{Luo et~al.(2012)Luo, Zhang, Zhao et~al.}]{1674-4527-12-9-004}
Luo, A.-L., Zhang, H.-T., Zhao, Y.-H., et~al. 2012, Research in Astronomy and
  Astrophysics, 12, 1243

\bibitem[{{Luo} et~al.(2015){Luo}, {Zhao}, {Zhao} et~al.}]{2015RAA....15.1095L}
{Luo}, A.-L., {Zhao}, Y.-H., {Zhao}, G., et~al. 2015, Research in Astronomy and
  Astrophysics, 15, 1095

\bibitem[{{Luridiana} et~al.(2015){Luridiana}, {Morisset}, \&
  {Shaw}}]{2015AA...573A..42L}
{Luridiana}, V., {Morisset}, C., \& {Shaw}, R.~A. 2015, \aap, 573, A42

\bibitem[{{Ly} et~al.(2016{\natexlab{a}}){Ly}, {Malhotra}, {Malkan}
  et~al.}]{2016ApJS..226....5L}
{Ly}, C., {Malhotra}, S., {Malkan}, M.~A., et~al. 2016{\natexlab{a}}, \apjs,
  226, 5

\bibitem[{{Ly} et~al.(2014){Ly}, {Malkan}, {Nagao}
  et~al.}]{2014ApJ...780..122L}
{Ly}, C., {Malkan}, M.~A., {Nagao}, T., et~al. 2014, \apj, 780, 122

\bibitem[{{Ly} et~al.(2016{\natexlab{b}}){Ly}, {Malkan}, {Rigby}, \&
  {Nagao}}]{2016ApJ...828...67L}
{Ly}, C., {Malkan}, M.~A., {Rigby}, J.~R., \& {Nagao}, T. 2016{\natexlab{b}},
  \apj, 828, 67

\bibitem[{{Ly} et~al.(2015){Ly}, {Rigby}, {Cooper}, \&
  {Yan}}]{2015ApJ...805...45L}
{Ly}, C., {Rigby}, J.~R., {Cooper}, M., \& {Yan}, R. 2015, \apj, 805, 45

\bibitem[{{Maiolino} et~al.(2008){Maiolino}, {Nagao}, {Grazian}
  et~al.}]{2008AA...488..463M}
{Maiolino}, R., {Nagao}, T., {Grazian}, A., et~al. 2008, \aap, 488, 463

\bibitem[{{Mannucci} et~al.(2010){Mannucci}, {Cresci}, {Maiolino}, {Marconi},
  \& {Gnerucci}}]{2010MNRAS.408.2115M}
{Mannucci}, F., {Cresci}, G., {Maiolino}, R., {Marconi}, A., \& {Gnerucci}, A.
  2010, \mnras, 408, 2115

\bibitem[{{Markwardt} et~al.(2009){Markwardt}, {Swank}, {Barthelmy}
  et~al.}]{2009ATel.2258....1M}
{Markwardt}, C.~B., {Swank}, J.~H., {Barthelmy}, S.~D., et~al. 2009, The
  Astronomer's Telegram, 2258, 1

\bibitem[{{Nicholls} et~al.(2013){Nicholls}, {Dopita}, {Sutherland}, {Kewley},
  \& {Palay}}]{2013ApJS..207...21N}
{Nicholls}, D.~C., {Dopita}, M.~A., {Sutherland}, R.~S., {Kewley}, L.~J., \&
  {Palay}, E. 2013, \apjs, 207, 21

\bibitem[{{Nicholls} et~al.(2014){Nicholls}, {Dopita}, {Sutherland}
  et~al.}]{2014ApJ...786..155N}
{Nicholls}, D.~C., {Dopita}, M.~A., {Sutherland}, R.~S., et~al. 2014, \apj,
  786, 155

\bibitem[{{Pustilnik} et~al.(2002){Pustilnik}, {Kniazev}, {Masegosa}
  et~al.}]{2002AA...389..779P}
{Pustilnik}, S.~A., {Kniazev}, A.~Y., {Masegosa}, J., et~al. 2002, \aap, 389,
  779

\bibitem[{{Pustilnik} \& {Martin}(2007)}]{2007AA...464..859P}
{Pustilnik}, S.~A., \& {Martin}, J.-M. 2007, \aap, 464, 859

\bibitem[{{Salim} et~al.(2014){Salim}, {Lee}, {Ly}
  et~al.}]{2014ApJ...797..126S}
{Salim}, S., {Lee}, J.~C., {Ly}, C., et~al. 2014, \apj, 797, 126

\bibitem[{{Shaw} et~al.(1998){Shaw}, {de La Pena}, {Katsanis}, \&
  {Williams}}]{1998ASPC..145..192S}
{Shaw}, R.~A., {de La Pena}, M.~D., {Katsanis}, R.~M., \& {Williams}, R.~E.
  1998, in Astronomical Data Analysis Software and Systems VII,
  \emph{Astronomical Society of the Pacific Conference Series}, vol. 145,
  edited by R.~{Albrecht}, R.~N. {Hook}, \& H.~A. {Bushouse}, 192

\bibitem[{{Shaw} \& {Dufour}(1995)}]{1995PASP..107..896S}
{Shaw}, R.~A., \& {Dufour}, R.~J. 1995, \pasp, 107, 896

\bibitem[{{Shi} et~al.(2005){Shi}, {Kong}, {Li}, \&
  {Cheng}}]{2005AA...437..849S}
{Shi}, F., {Kong}, X., {Li}, C., \& {Cheng}, F.~Z. 2005, \aap, 437, 849

\bibitem[{{Skillman} et~al.(1989){Skillman}, {Kennicutt}, \&
  {Hodge}}]{1989ApJ...347..875S}
{Skillman}, E.~D., {Kennicutt}, R.~C., \& {Hodge}, P.~W. 1989, \apj, 347, 875

\bibitem[{{Skrutskie} et~al.(2006){Skrutskie}, {Cutri}, {Stiening}
  et~al.}]{2006AJ....131.1163S}
{Skrutskie}, M.~F., {Cutri}, R.~M., {Stiening}, R., et~al. 2006, \aj, 131, 1163

\bibitem[{{Song} et~al.(2012){Song}, {Luo}, {Comte}
  et~al.}]{2012RAA....12..453S}
{Song}, Y.-H., {Luo}, A.-L., {Comte}, G., et~al. 2012, Research in Astronomy
  and Astrophysics, 12, 453

\bibitem[{{Storey} et~al.(2014){Storey}, {Sochi}, \&
  {Badnell}}]{2014MNRAS.441.3028S}
{Storey}, P.~J., {Sochi}, T., \& {Badnell}, N.~R. 2014, \mnras, 441, 3028

\bibitem[{Su \& Cui(2004)}]{1009-9271-4-1-1}
Su, D.-Q., \& Cui, X.-Q. 2004, Chinese Journal of Astronomy and Astrophysics,
  4, 1

\bibitem[{{Tayal} \& {Zatsarinny}(2010)}]{2010ApJS..188...32T}
{Tayal}, S.~S., \& {Zatsarinny}, O. 2010, \apjs, 188, 32

\bibitem[{{Tremonti} et~al.(2004){Tremonti}, {Heckman}, {Kauffmann}
  et~al.}]{2004ApJ...613..898T}
{Tremonti}, C.~A., {Heckman}, T.~M., {Kauffmann}, G., et~al. 2004, \apj, 613,
  898

\bibitem[{{Veilleux} \& {Osterbrock}(1987)}]{1987ApJS...63..295V}
{Veilleux}, S., \& {Osterbrock}, D.~E. 1987, \apjs, 63, 295

\bibitem[{Wang et~al.(1996)Wang, Su, Chu, Cui, \& Wang}]{Wang:96}
Wang, S., Su, D., Chu, Y., Cui, X., \& Wang, Y. 1996, Appl. Opt., 35, 5155

\bibitem[{{Wright} et~al.(2010){Wright}, {Eisenhardt}, {Mainzer}
  et~al.}]{2010AJ....140.1868W}
{Wright}, E.~L., {Eisenhardt}, P.~R.~M., {Mainzer}, A.~K., et~al. 2010, \aj,
  140, 1868-1881

\bibitem[{{Yates} et~al.(2012){Yates}, {Kauffmann}, \&
  {Guo}}]{2012MNRAS.422..215Y}
{Yates}, R.~M., {Kauffmann}, G., \& {Guo}, Q. 2012, \mnras, 422, 215

\bibitem[{{Zaritsky} et~al.(1994){Zaritsky}, {Kennicutt}, \&
  {Huchra}}]{1994ApJ...420...87Z}
{Zaritsky}, D., {Kennicutt}, R.~C., Jr., \& {Huchra}, J.~P. 1994, \apj, 420, 87

\bibitem[{{Zhao} et~al.(2012){Zhao}, {Zhao}, {Chu}, {Jing}, \&
  {Deng}}]{2012RAA....12..723Z}
{Zhao}, G., {Zhao}, Y.-H., {Chu}, Y.-Q., {Jing}, Y.-P., \& {Deng}, L.-C. 2012,
  Research in Astronomy and Astrophysics, 12, 723

\end{thebibliography}
